\documentclass[%
 reprint,
 amsmath,amssymb,
 aps,
floatfix,
]{revtex4-2}

\usepackage{float}
\usepackage{subcaption}
\usepackage{graphicx}
\usepackage{dcolumn}
\usepackage{bm}
\usepackage[breaklinks=true]{hyperref}
\newcommand{\comment}[1]{} 

\begin{document}

\preprint{APS/123-QED}

\title[]{Neutrino Energy Estimation in Neutral Current Interactions and \\Prospects for Sterile Neutrino Searches}

\author{A. P. Furmanski}
 \email{afurmans@umn.edu}
 \affiliation{School of Physics and Astronomy, University of Minnesota, Minneapolis, MN 55455, USA }%
\author{C. Hilgenberg}
 \email{chilgenb@umn.edu.}
 \affiliation{School of Physics and Astronomy, University of Minnesota, Minneapolis, MN 55455, USA }%

\date{\today}

\begin{abstract}
Modern neutrino detectors, particularly the large liquid argon time projection chambers (LAr TPCs) at SBN and DUNE, provide an unprecedented amount of information about GeV-scale interactions.
By taking advantage of the excellent spatial and calorimetric resolution as well as the low tracking thresholds provided by LAr TPCs, we present a novel method of estimating the neutrino energy in neutral current interactions.
This method has potential implications for the search for a sterile neutrino; it allows for the potential observation of spectral distortions due to sterile neutrino-induced oscillations in the neutral current neutrino energy spectrum.
As an example, we use this method to perform an analysis of the statistics-only sensitivity to sterile neutrinos in the neutral current channel at SBN under a 3+1 model.
\end{abstract}

\maketitle

\section{Introduction} 
Accurately and precisely reconstructing the incident neutrino energy in neutrino interactions is notoriously difficult. The level of difficulty depends on whether the neutrino interaction is charged current (CC) or neutral current (NC). CC interactions provide a means of identifying the scattering neutrino's flavor. In addition, in the few-GeV neutrino energy regime where quasielastic (QE) scattering is the dominant interaction mode, the kinematics of the outgoing lepton are sufficient for determining the neutrino energy. This gets more challenging at higher energies where resonant production and deep inelastic scattering dominate and final state interactions (FSI) are more prevalent. Nevertheless, the resolution on the neutrino energy is typically at the level of tens of percent. 

Neutral current interactions provide neither the neutrino flavor nor knowledge of the energy carried away by the outgoing lepton. The only handle on the neutrino energy comes from the final state hadronic system. 

The typical method employed in estimating the incident neutrino energy in NC interactions is to rely on calorimetry alone, summing all visible energy in the detector to provide a lower bound. This method yields a poor energy resolution relative to CC interactions while also introducing a significant bias toward lower energies. The method's performance depends on the detector tracking thresholds and calorimetric resolution. 

Compared to water Cherenkov and scintillator based detectors, liquid argon time projection chambers (LAr TPCs) provide low tracking thresholds and precise energy and angular resolution for charged particles. This detector technology has been adopted by the Short-Baseline Neutrino Program (SBN)~\cite{acciarri2015proposal} and the future Deep Underground Neutrino Experiment (DUNE)~\cite{dune_tdr_v1}\cite{dune_tdr_v2}\cite{dune_tdr_v3}\cite{dune_tdr_v4}. LAr TPCs, in combination with intense accelerator based neutrino sources, are providing and will provide unprecedented detail about neutrino interactions in the few-GeV neutrino energy regime.

With the unprecedented level of detail about the hadronic final state provided by LAr TPCs, it might be possible to improve neutrino energy estimation in NC interactions. In the first half of this paper, we revisit the visible energy based method and introduce a new method based on the final state hadronic system kinematics. We present a toy analysis that demonstrates the proof of concept of our kinematic method. 

In the second half of this paper, we explore one possible application of neutral current interactions reconstructed with the kinematic method: the sterile neutrino search at SBN. If the incident neutrino energy can be reconstructed sufficiently well, NC interactions can be used to probe neutrino oscillation parameters that are challenging or impossible to probe with CC channels. We present a statistics-only toy analysis of NC disappearance sensitivities, illustrating that the addition of our reconstruction method has the potential to improve sensitivity.  

\begin{figure}[th]
    \centering
    \includegraphics[width=0.4\linewidth]{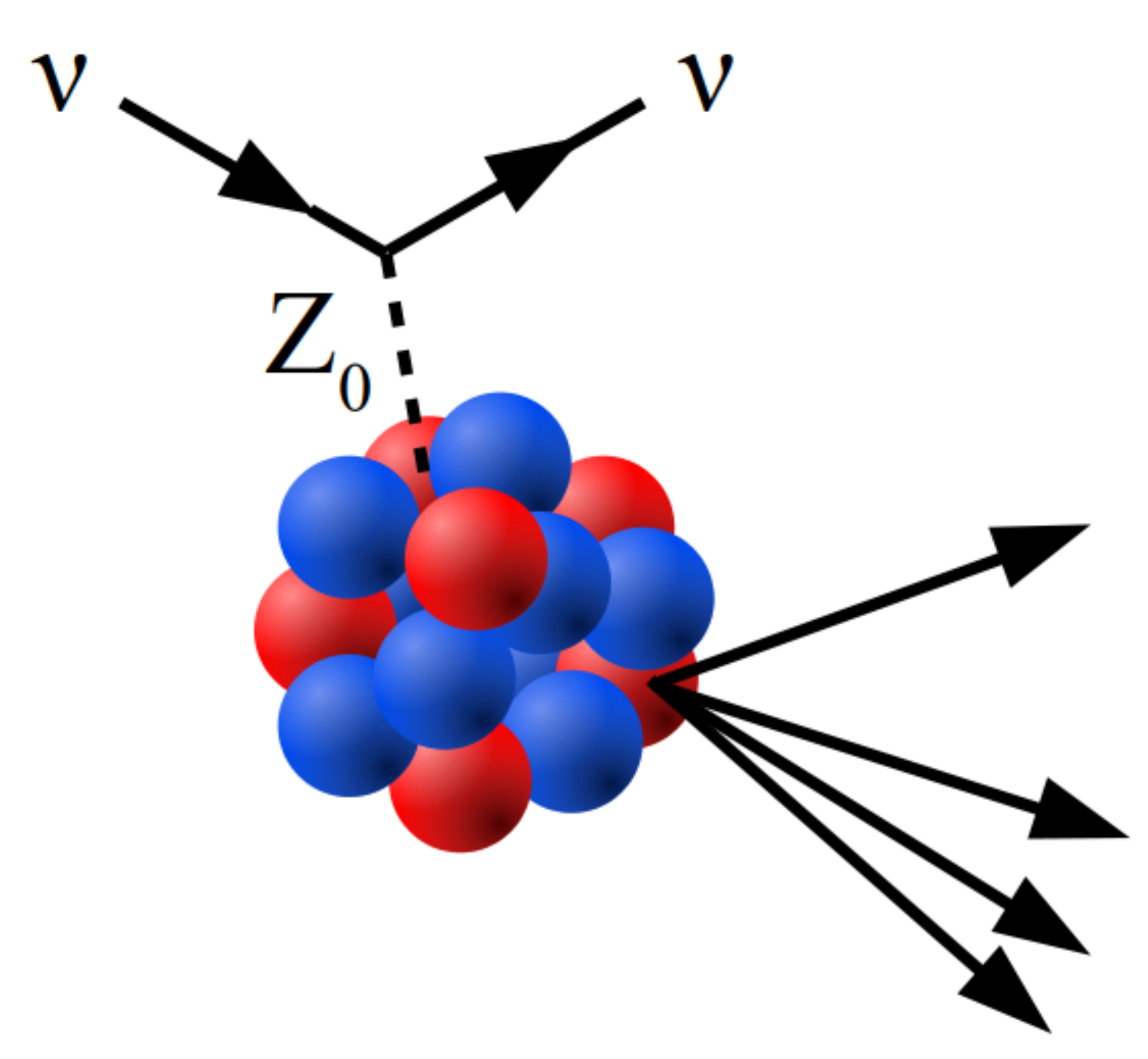}
    \caption{A neutrino scatters off a nucleon, producing a final state hadronic system that can be used to determine the incident neutrino energy.}
    \label{fig:nc_scatter}
\end{figure}

\section{Neutrino Energy Reconstruction in Neutral Current Interactions}

In NC interactions, the scattering neutrino carries away some fraction of its initial energy as illustrated in Figure~\ref{fig:nc_scatter}. A trivial 
lower bound on the neutrino energy can be obtained by summing over all visible deposited energy associated with a neutrino vertex candidate. This calorimetric method leads to a significant underestimation of the neutrino energy.  

Significant improvements in neutral current neutrino energy reconstruction would be beneficial for several different kinds of measurements: improved characterization of NC backgrounds, more precise measurements of NC cross sections, gleaning shape information for oscillation searches, etc. This motivates the search for alternatives to the conventional calorimetric method. In what follows, we present a comparison of the conventional calorimetric method and a new reconstruction method based on the final state hadronic system kinematics.

We evaluate reconstruction performance using a sample of five million $\nu_\mu$ and $\bar{\nu}_\mu$ interactions on argon, each generated using GENIE v3.00.06~\cite{Andreopoulos:2009rq} with the G18\_10a configuration and the G18\_10a\_02\_11a tune. This configuration includes a local Fermi gas nuclear model and an empirical meson exchange current model. 

For this study, we use the SBN Booster Neutrino Beam (BNB) neutrino flux, taken from ~\cite{acciarri2015proposal}. The flux is broad-band, peaking at 600 MeV (see Figure~\ref{fig:flux}). The flux prediction for the neutrino running configuration (forward horn current) at each SBN detector is used to generate detector-specific, SBN-like samples. A combination of the flux and detector active mass is used to generate a normalization factor for each flux component in terms of protons on target (POT). Detector baselines, active masses, and POT are provided in Section~III.
\subsection{Neutrino Energy from Visible Energy}
Relying on calorimetry alone, to measure the total energy deposited in a detector following a NC interaction, does not account for the energy carried away by the scattering neutrino. This feature combined with threshold effects leads to a significant underestimation of the initial neutrino energy. In our case, this underestimation is at the level of 90\%. This bias can be simulated, and the effect can be corrected, usually by using Monte Carlo templates or unfolding techniques. However, model based corrections cannot improve poor energy resolution. Furthermore, reliance on such corrections can introduce systematic uncertainties into the analysis. Despite these limitations, this calorimetric method has been used with success, by MINOS/MINOS+\cite{minossterile} and NO$\nu$A~\cite{nova_thesis}\cite{nova_ncdis} for example. 

\begin{figure}[htb]
    \centering
    \begin{subfigure}{\linewidth}
        \includegraphics[width=\linewidth]{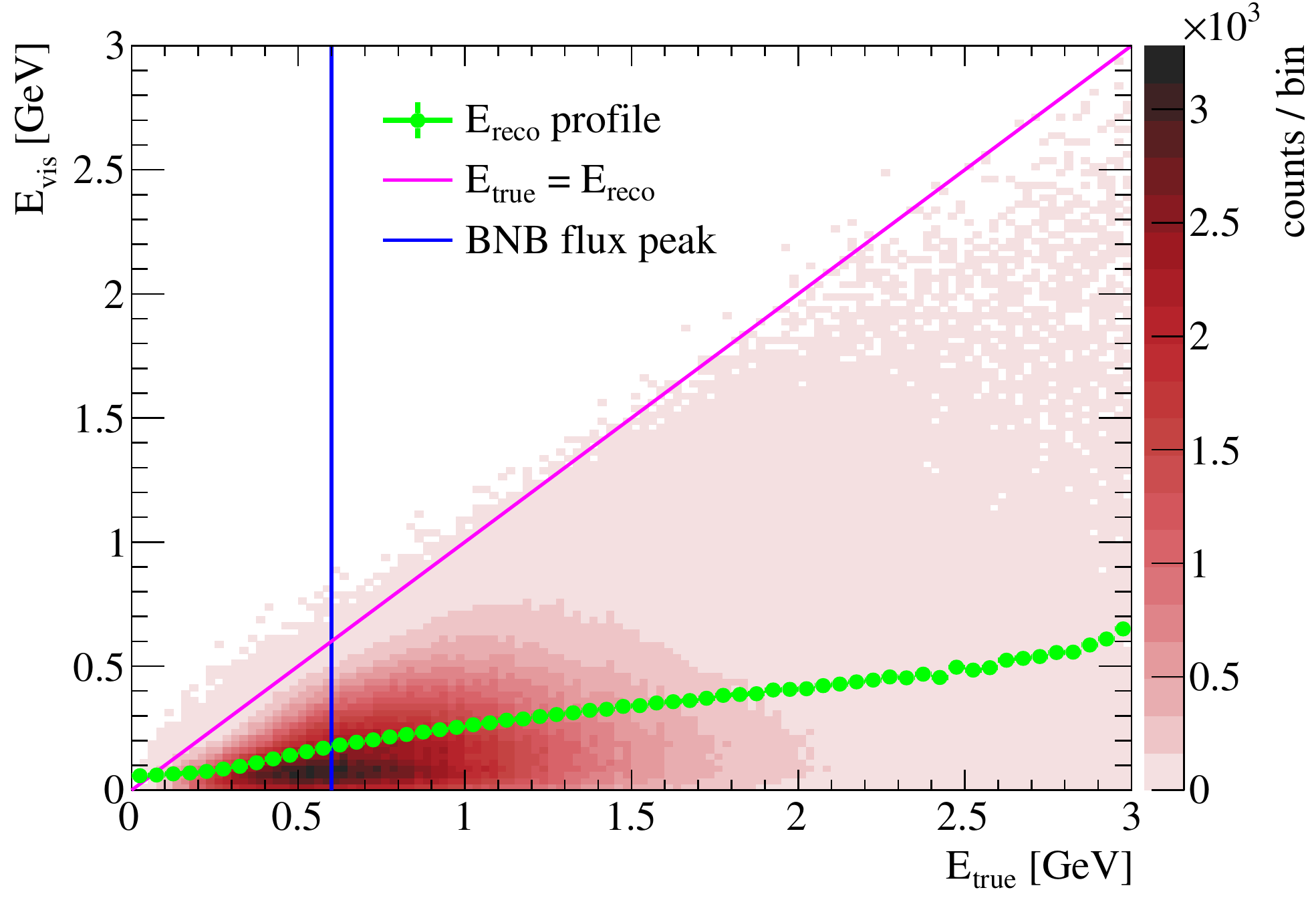}
        \subcaption{No correction.}
        \label{fig:enu_from_vis}
    \end{subfigure} %
    \begin{subfigure}{\linewidth}   
        \includegraphics[width=\linewidth]{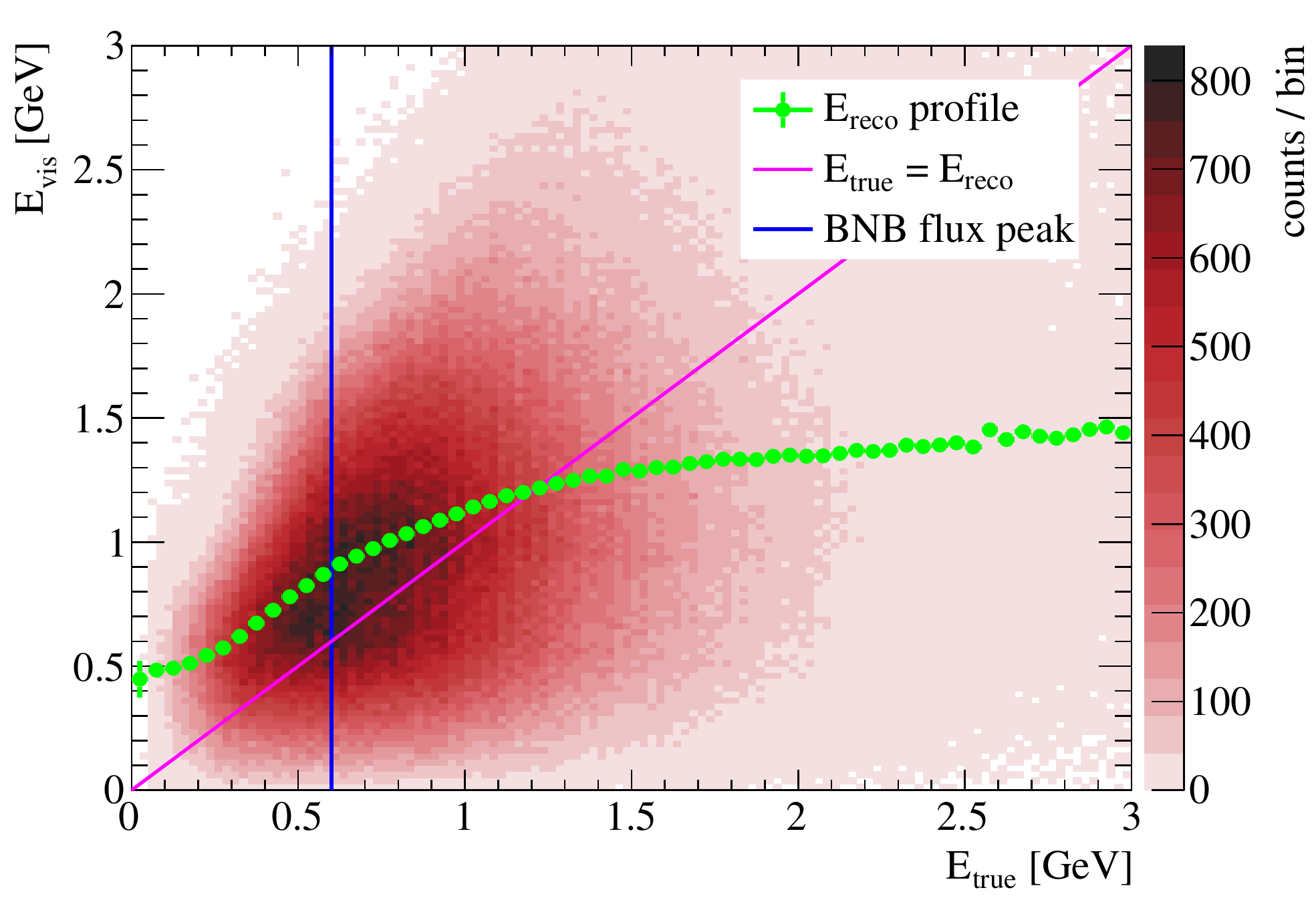}  
        \subcaption{With correction.}
        \label{fig:enu_from_vis_corr}
    \end{subfigure}
    \caption{Visible hadronic energy (y-axis) in NC0$\pi^{+/-}$ interactions vs. true neutrino energy (x-axis) shows that the calorimetric method significantly underestimates the neutrino energy (a). A correction derived from Monte Carlo can remove the bias (as shown by the reconstructed energy profile), but it cannot improve the resolution (b).}
\end{figure}

The initial neutrino energy, reconstructed with the calorimetric method, is plotted against the true neutrino energy in Figure~\ref{fig:enu_from_vis} for a NC0$\pi^{+/-}$ selection; the reason for this choice will become clear in Section~III. The reconstructed energy includes approximate detector effects, discussed later. The impact of the missing energy is evident. The mean reconstructed energy profile is also shown, approximately linear in true energy. 

In order to make a fair comparison between the calorimetric and the kinematic method, discussed next, we attempt to correct the reconstructed energy. A fit to the ratio of the true to the reconstructed neutrino energy as a function of reconstructed energy is used to calculate a correction on an event-by-event basis.
By construction, this correction makes the median bias zero.
The corrected version of Figure~\ref{fig:enu_from_vis} is shown in Figure~\ref{fig:enu_from_vis_corr}.
Profiles based on the mean show the impact of the highly asymmetric resolution, leading to large biases remaining in the mean corrected energy.
Additionally, while a correction can significantly reduce the bias, the resolution remains poor.

\subsection{Neutrino Energy from Final State Kinematics}

The capabilities of LAr TPCs motivate a new reconstruction method 
based on the kinematics of the final state hadronic system. We have developed a kinematic model that, using some simplifying assumptions, uniquely determines the initial neutrino energy, accounting for the energy carried away by the neutrino. We assume that the neutrino scatters off a single nucleon at rest. If we ignore nuclear effects, including nucleon-nucleon correlation, nuclear recoil, and binding energy, the incoming neutrino energy can be determined from the energy and momentum of the final state hadronic system.
\begin{figure}[htb]
    \centering
    \begin{subfigure}{\linewidth}
      \centering
      \includegraphics[width=\linewidth]{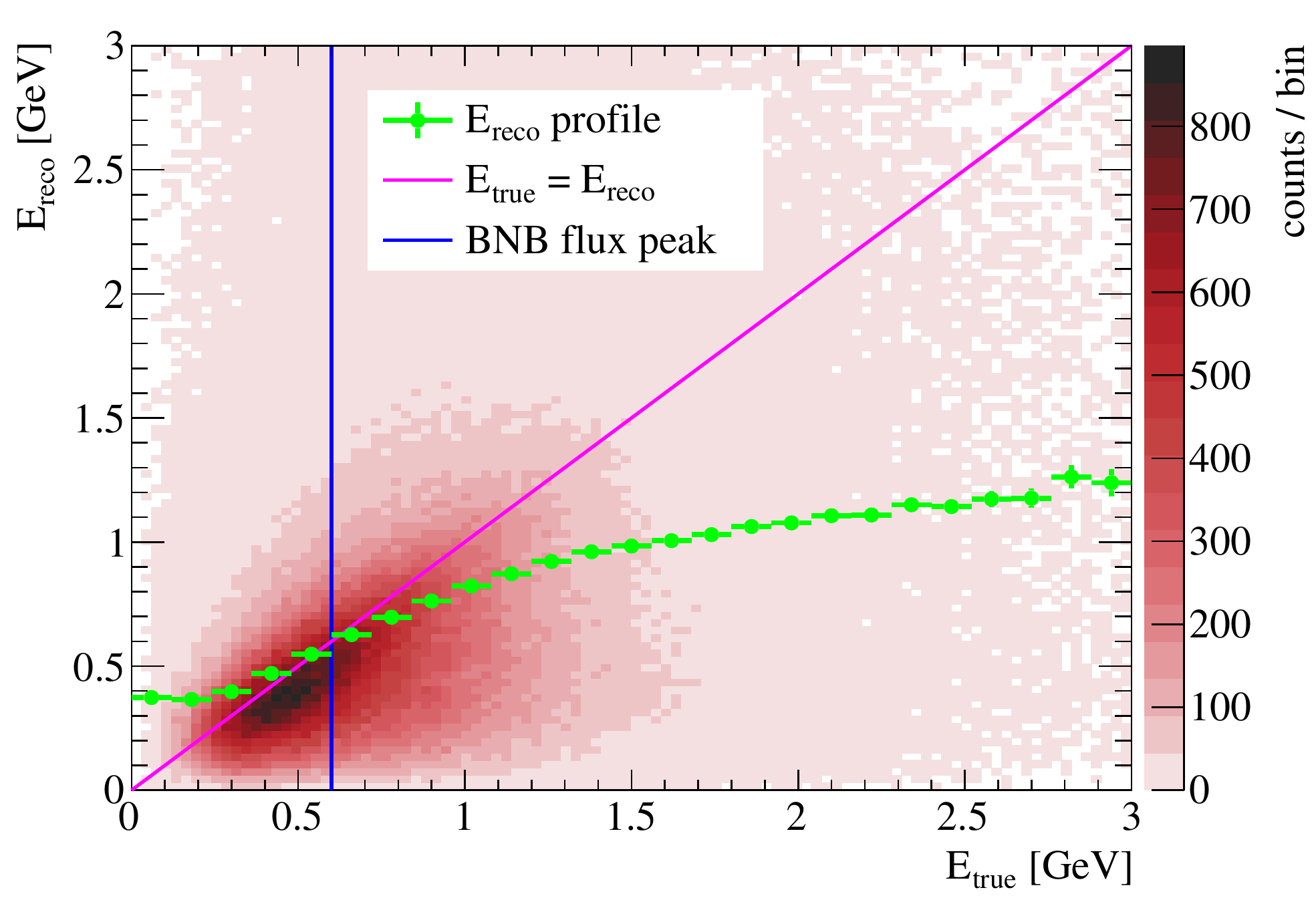}
      \subcaption{No correction.}
      \label{fig:kin_2d}  
    \end{subfigure} %
    \begin{subfigure}{\linewidth}
      \centering
      \includegraphics[width=\linewidth]{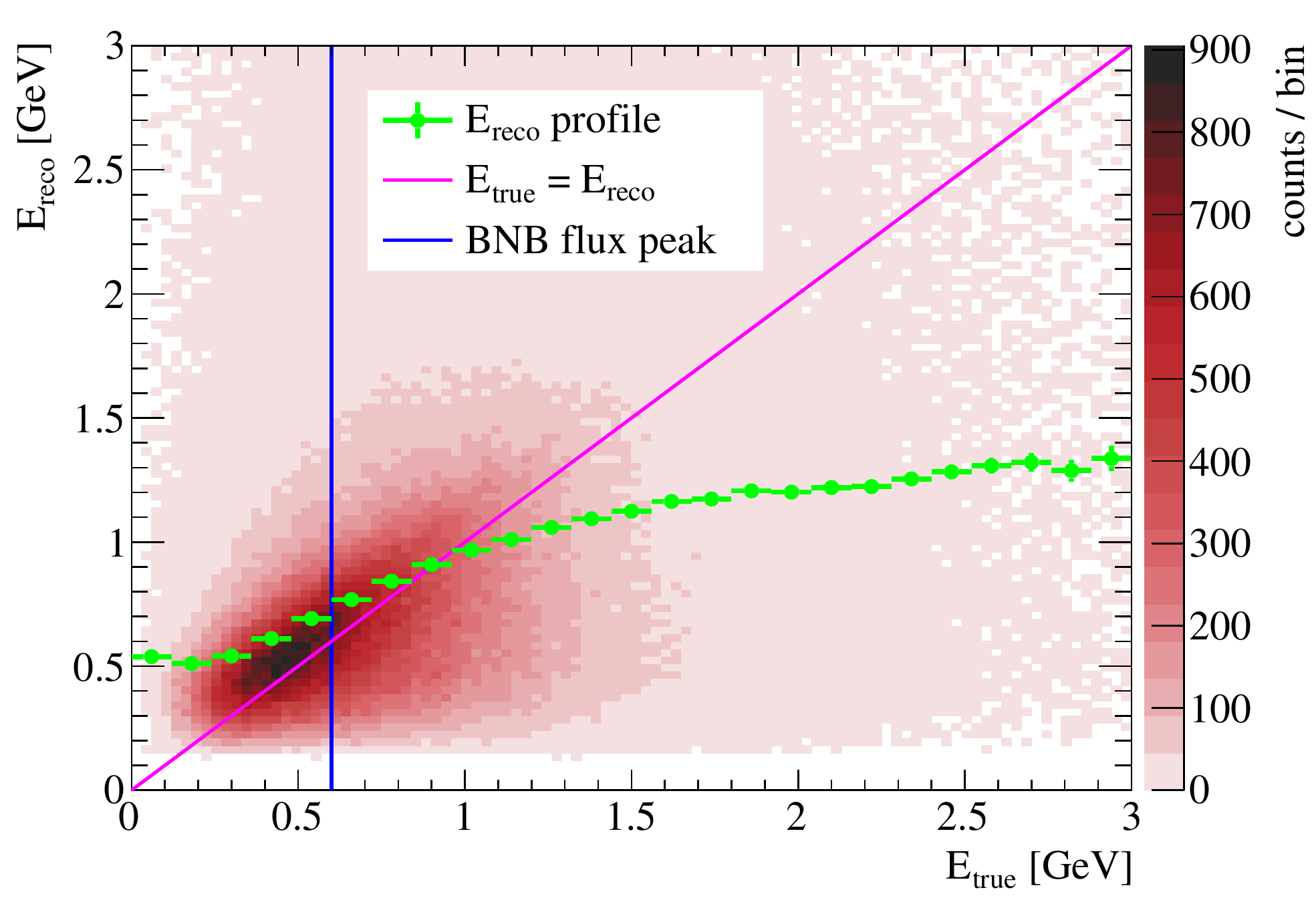}
      \subcaption{With correction.}      
      \label{fig:kin_2d_corr}
    \end{subfigure} 
    \caption{Reconstructed neutrino energy using the kinematic method (y-axis) vs. true neutrino energy (x-axis) for NC1p events with neutron tagging shows reasonable agreement, peaking along the diagonal (a). An overall scale correction is applied to remove bias (b). The reconstructed energy profile shows minimal bias at or around the BNB flux peak.}
\end{figure}

In the lab frame, the initial state neutrino and nucleon four momenta are given by $p_{\nu,i}^\mu = (E_\nu,0,0,p_\nu)$ and $p_{N,i}^\mu = (m_N,0,0,0)$ with $m_N$ taken to be the average of the neutron and proton masses. The z-axis is aligned with the neutrino beam direction. In the final state, also in the lab frame, the outgoing neutrino is unmeasured, but the four momentum of the hadronic system can be described by $p_h^\mu = (E_h,\boldsymbol{p_h})$. From this, we calculate the initial neutrino energy in terms of the final state hadronic system kinematics, shown in Equation~\ref{eq:enu} where $\theta_h$ is the hadronic system angle with respect to the beam direction ($\cos\theta_h = \mathbf{p_h} \cdot \hat{z} / |\mathbf{p_h}|$). The hadronic four-momentum is calculated by summing the four-momenta of all visible particles. This method is analogous to the CC energy reconstruction used by T2K and MiniBooNE, based on the quasi-elastic assumption when only the final state lepton is observed.

\begin{equation}
    \label{eq:enu}
    E_\nu^{\text{reco}} = \frac{p_h^2-(E_h-m_N)^2}{2(m_N+p_h \text{cos}\theta_h-E_h)}
\end{equation}

Figure~\ref{fig:kin_2d}, with the reconstructed neutrino energy plotted as a function of the true neutrino energy, demonstrates the validity of our method. The figure was obtained from a NC1p selection with neutron tagging, discussed in the next subsection. In contrast to the calorimetric method, the distribution obtained with our method is peaked along the diagonal without any Monte Carlo corrections. For consistency, as we apply a correction to the kinematic method, we apply a small scale correction to remove any bias. The correction is obtained by fitting the absolute energy resolution with a Gaussian to extract the peak value.
The mean reconstructed energy does still show a bias, as in the calorimetric case, but the resolution is significantly better.

Having introduced both reconstruction methods, we now discuss how we account for  detector effects in our analysis.

\subsection{Adding Detector Effects}

For simplicity, we forgo a full detector simulation and assume Gaussian angular and energy resolutions for the final state particles. Table~\ref{tab:performance} summarizes a reasonable range of resolutions by particle type. In addition, the tracking thresholds, applied to the true kinetic energy, are provided. The ranges of possible reconstruction performances are taken from published ArgoNeut~\cite{PhysRevD.89.112003} and MicroBooNE analyses~\cite{Abratenko2020}\cite{Abratenko2020_2}\cite{Abratenko2019}\cite{Adams_2020}.
\begin{table}[htb]
    \caption{Assumed performances for LAr TPCs are given for species of interest. Charged pions with momenta below 300~MeV/c are assumed to have similar reconstruction performances as muons. The threshold is set by the requirement that two or more TPC wires are crossed and is applied to the true kinetic energy. }
    \centering
    \begin{tabular}{c|c|c|c}
        Species & Threshold & Energy     & Angular \\  
                & [MeV]     & Resolution & Resolution [deg] \\
         \hline
         p \cite{PhysRevD.89.112003}\cite{Abratenko2020} & 25-50 & 60 MeV & 5-10 \\
         $\pi^{+/-}$ \cite{Abratenko2020_2}\cite{Abratenko2019}& 10-20 & 10-20\% & 2-5 \\
         $\gamma$ \cite{Adams_2020}& 30 & 10-20\% & 5-10 \\
    \end{tabular}
    \label{tab:performance}
\end{table}

If particle species not listed in Table~\ref{tab:performance} are present in a given NC event, with the exception of neutrons or neutral pions, the event is excluded from the analysis. This results in a negligible loss in sample size while making the interpretation of our results more straightforward. Neutral pions are not reconstructed directly, by calculating the invariant mass of the decay photons for example; only the decay photons are used. 

In comparing reconstruction performances between the calorimetric and kinematic methods, discussed next, we adopt the most optimistic choice of the performances listed in Table~\ref{tab:performance}.

\subsection{Evaluating Reconstruction Performance}

In order to evaluate the reconstruction performance of each reconstruction method, we first compare the neutrino energy resolutions. Next, we compare the reconstruction efficiencies. In both cases, we consider different sample selections. We adopt a NC0$\pi^{+/-}$ selection for the calorimetric method.
\begin{figure}[htb]
    \centering
    \begin{subfigure}{\linewidth}
        \includegraphics[width=\linewidth]{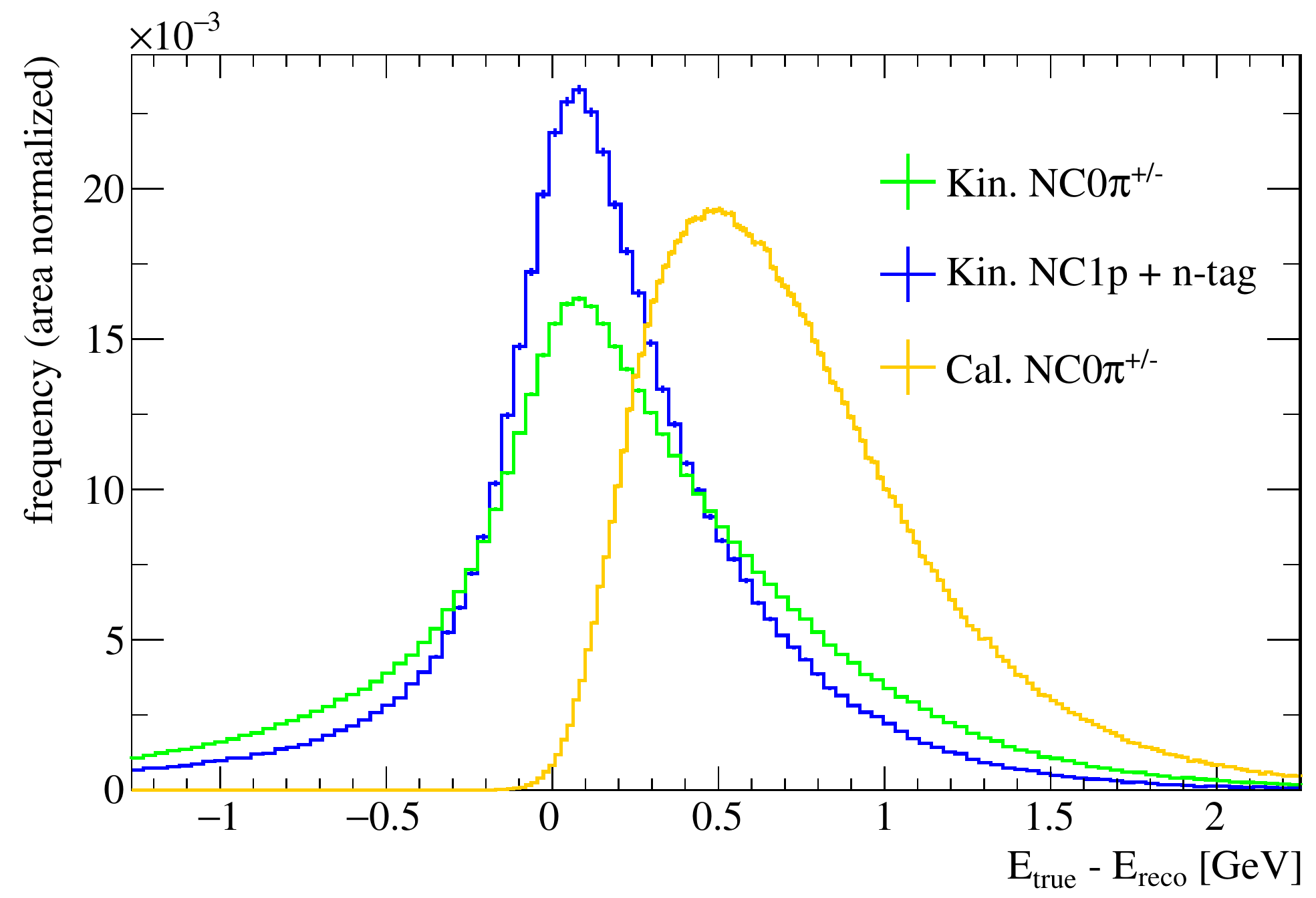}
        \subcaption{No correction.}
        \label{fig:diff}
    \end{subfigure}
    \begin{subfigure}{\linewidth}
        \includegraphics[width=\linewidth]{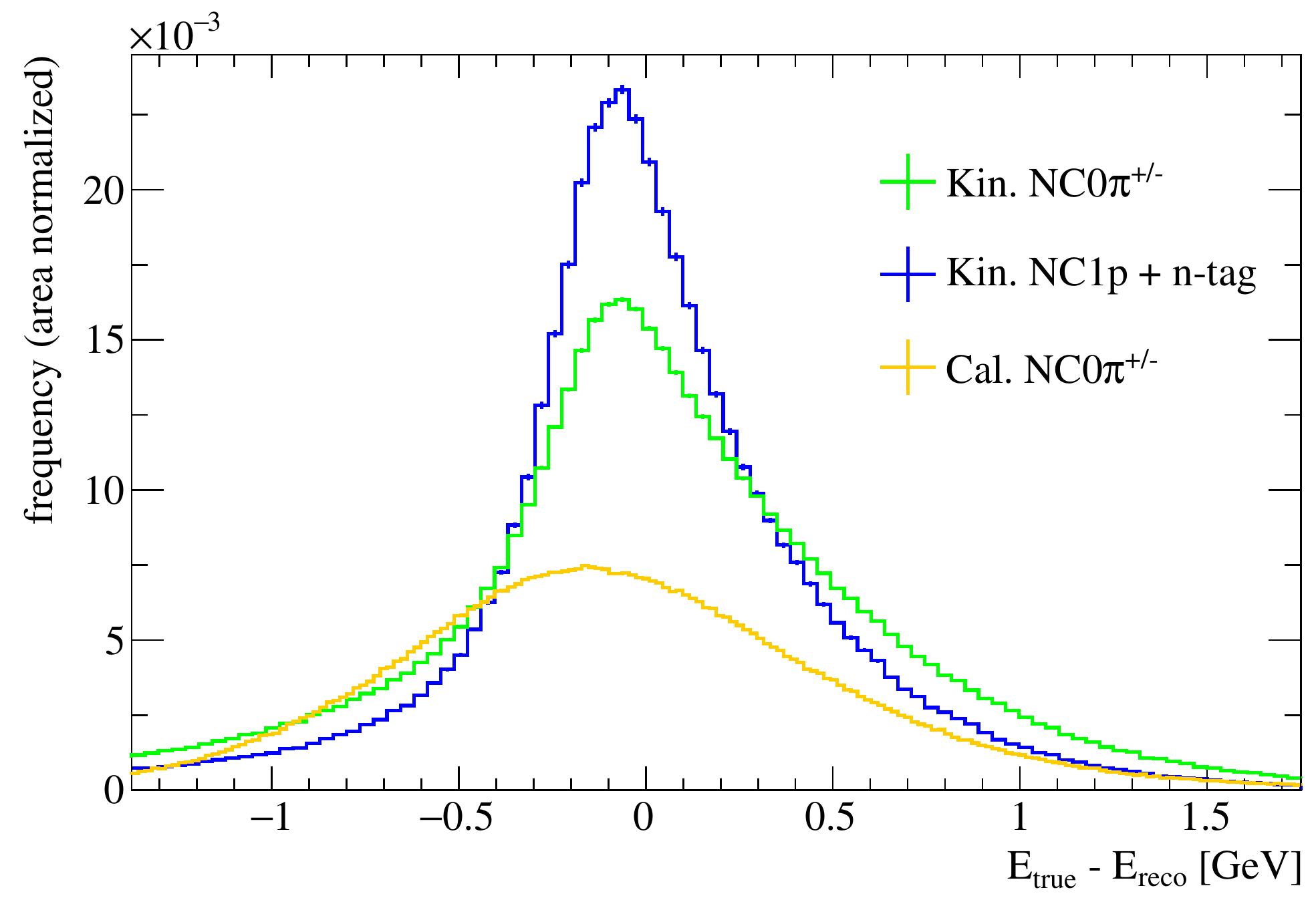}
        \subcaption{With correction.}
        \label{fig:diff_corr}
    \end{subfigure}   
    \caption{Absolute energy resolution obtained with different reconstruction methods shows that the kinematic method has better energy resolution and lower bias than the calorimetric method. Each distribution has been area normalized. Results for no Monte Carlo correction (a) and with the correction (b) show that, while the correction removes bias, it does not improve resolution.}
    \label{fig:diffs}
\end{figure}
 For the kinematic method, we investigated several exclusive final state topologies. Here, we show results for NC0$\pi^{+/-}$ as well as NC1p, the dominant topology at BNB energies. The other topologies we investigated were statistically limited. While other topologies are not considered here, it is worth noting that if the kinematic method were to be applied at DUNE, where the neutrino flux peaks in the few-GeV range, other final state topologies may be relevant.
\begin{figure}[htb]
    \centering
    \begin{subfigure}{\linewidth}
        \includegraphics[width=\linewidth]{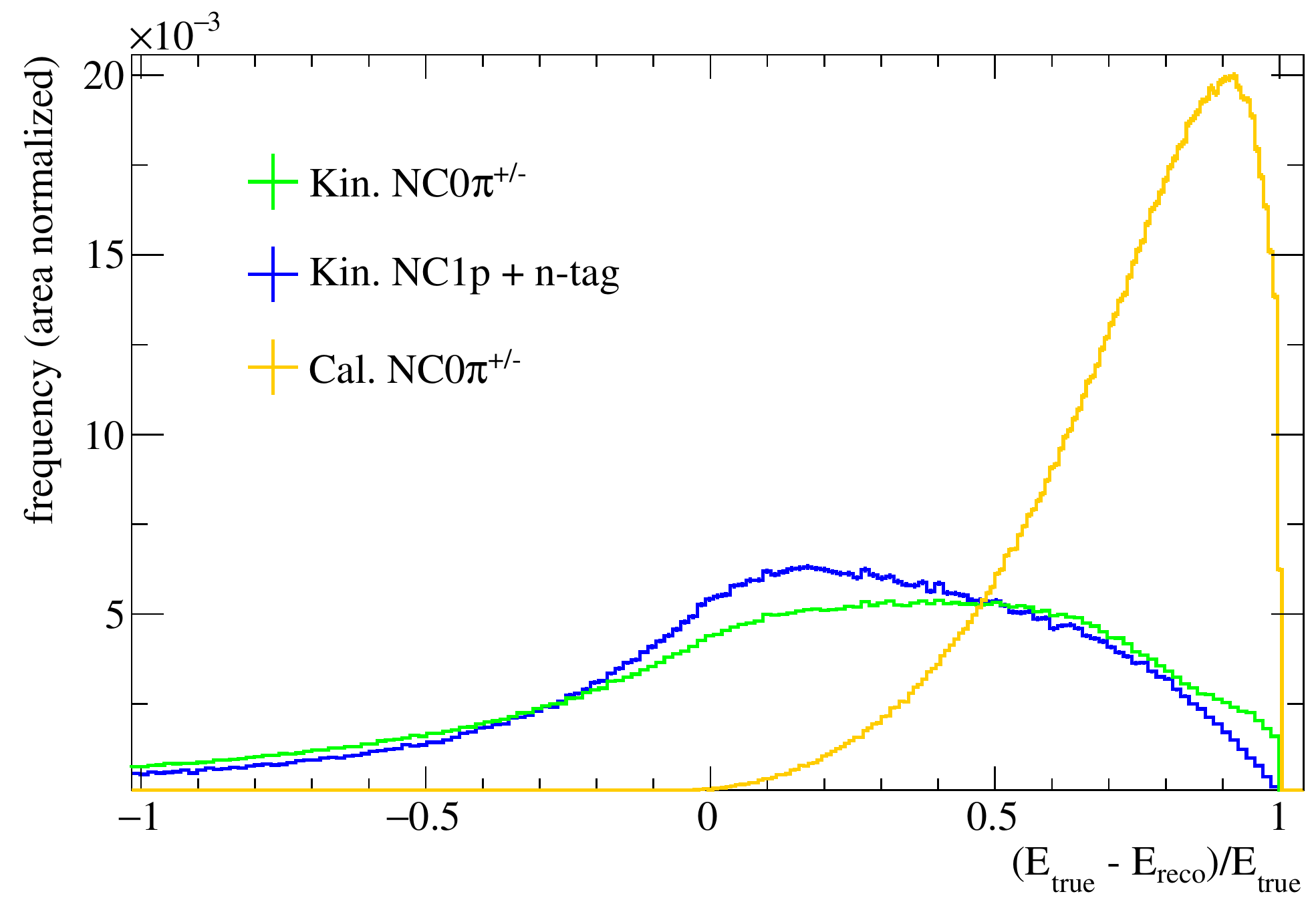}
        \subcaption{No correction.}
        \label{fig:frac}
    \end{subfigure}
    \begin{subfigure}{\linewidth}
        \includegraphics[width=\linewidth]{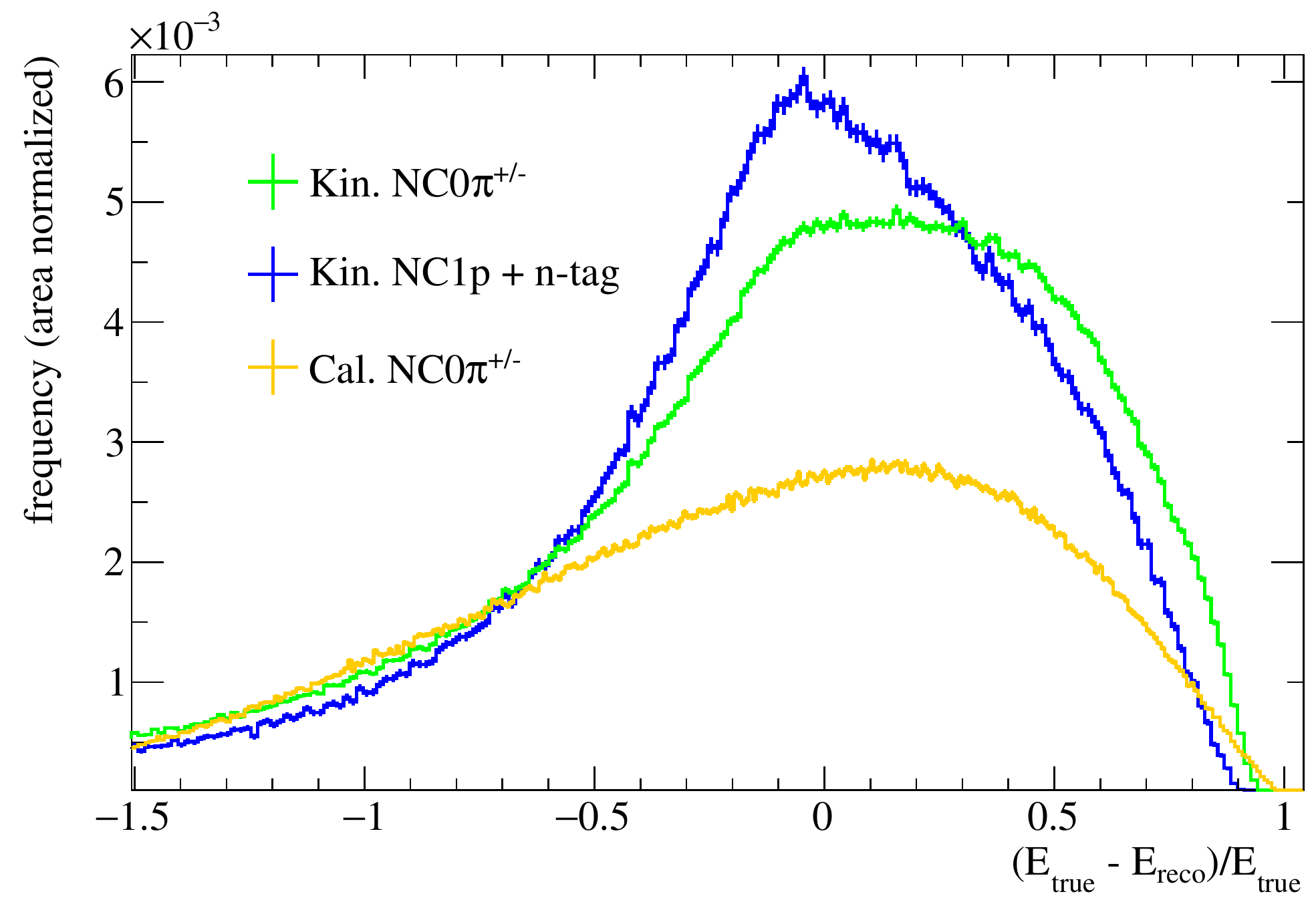}
        \subcaption{With correction.}
        \label{fig:frac_corr}
    \end{subfigure}
    \caption{Fractional energy resolution obtained with different reconstruction methods shows that the kinematic method has better energy resolution ($\sim$20\%) than the calorimetric method ($\sim$90\%). Each distribution has been area normalized. Results for no Monte Carlo correction (a) and with the correction (b) show that, while the correction removes bias, it does not improve resolution.}
    \label{fig:fracs}
\end{figure}

One of the most significant drivers for underestimating the incident neutrino energy in CC or NC interactions is neutrons. LAr TPCs have some ability to observe neutrons either through deexcitation photons produced in neutron capture or through the production of charged hadrons in inelastic scatters. We neglect deexcitation photons and conservatively estimate that we can tag neutrons (n-tagging) having kinetic energies above 50 MeV via inelastic scatters with 50\% efficiency. If one or more neutrons are tagged, we make no attempt to estimate the energy of the scattering neutron(s) and reject the whole event. This cut has been applied for the kinematic method with NC1p selection only.

The absolute reconstructed energy resolution is plotted in Figure~\ref{fig:diffs} while the fractional reconstructed resolution is plotted in Figure~\ref{fig:fracs}. Resolutions are shown for calorimetric NC0$\pi^{+/-}$, kinematic NC0$\pi^{+/-}$, and kinematic NC1p with n-tagging methods and selections. From both figures, it is evident that the kinematic method provides a less biased estimate and better resolution ($\sim$20\% vs. $\sim$90\%) of the neutrino energy. While Monte Carlo corrections are not necessary for the kinematic method, a desirable feature to be sure, a more fair comparison is between the Monte Carlo-corrected resolutions. With or without corrections, the energy resolution obtained by the kinematic method is significantly better.

Table~\ref{tab:res} summarizes the energy resolution and bias associated with each reconstruction method and sample selection discussed above. These results quantitatively demonstrate that the kinematic method provides 80\% lower bias and 25\% lower full width at half maximum (FWHM) compared to the calorimetric method without any corrections. With corrections, the kinematic method provides 75\% lower bias and 45\% lower FWHM.
\begin{table}[htb]
    \caption{ The neutrino energy bias and resolution is summarized for the two different reconstruction methods with different sample selections. Values are shown for both uncorrected (inside parentheses) and corrected (outside parentheses) cases.}
    \centering
    \begin{tabular}{c|c|c}
        Method and Selection & Bias [MeV] & FWHM [MeV] \\
         \hline
         Calorimetric NC0$\pi^{+/-}$ & 108 (662) & 1198 (824) \\
         Kinematic NC0$\pi^{+/-}$    & 20 (133)  & 748 (711) \\
         Kinematic NC1p + n-tag      & 34 (109)  & 524 (524) \\
    \end{tabular}
    \label{tab:res}
\end{table}

\begin{figure}[htb]
    \centering
    \includegraphics[width=\linewidth]{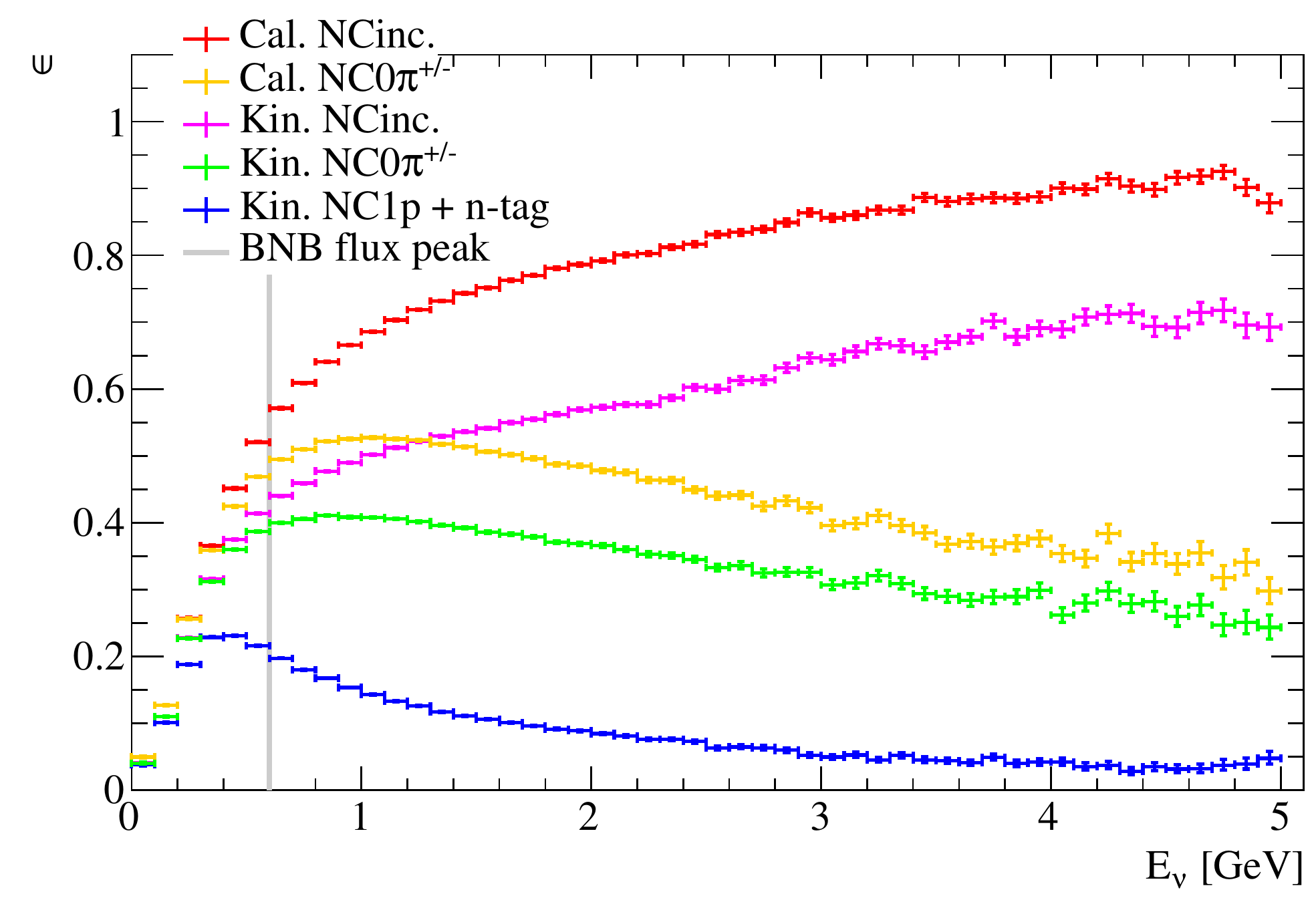}
    \caption{The reconstruction efficiency (y-axis) is plotted as a function of the true neutrino energy (x-axis) for the two different reconstruction methods, calorimetric or kinematic, and different sample selections: NC inclusive, NC0$\pi^{+/-}$, or NC1p with neutron tagging. The efficiency values are most significant near the BNB flux peak at $\sim$600~MeV.}
    \label{fig:eff}
\end{figure}

While the kinematic method provides superior neutrino energy resolution compared to the conventional calorimetric method, it incurs inefficiencies when the assumptions of the model fail. 
The model assumptions are valid primarily for NC quasi-elastic scattering, the dominant process at BNB energies, as evidenced by the best neutrino energy resolution being obtained with the NC1p with n-tagging selection. However, sub-leading processes are significant such as meson exchange current, resonance production, and deep inelastic scattering.

There is a long tail in the neutrino energy fractional resolution distribution where the energy is overestimated. This is primarily driven by the waning validity or outright failure of our model assumptions, the nucleon-nucleon correlations and the nucleon Fermi momentum in particular. In some cases, invalid model assumptions result in a negative reconstructed neutrino energy; these events are rejected, resulting in reduced efficiency with respect to the calorimetric method.

The energy dependence of the reconstruction efficiency for both methods with different sample selections is shown in Figure~\ref{fig:eff}. Here, the efficiency is calculated relative to the true number of NC inclusive events. The efficiency values are most important near the BNB flux peak.

For NC inclusive selections, the kinematic method has an approximately 15\% lower integrated efficiency. It should be noted, however, that LAr TPC detectors at these energies struggle to differentiate between muons and charged pions, each having nearly identical energy loss as a function of range. It is not unreasonable to assume, therefore, that to eliminate  $\nu_\mu$CC background contributions, events with charged pions in the final state may need to be rejected. This motivates the choice of the NC0$\pi^{+/-}$ selection.
\begin{table}[htb]
    \caption{Efficiencies integrated over the full range of neutrino energy are provided for the different reconstruction methods, calorimetric or kinematic, and event selections: NC inclusive, NC0$\pi^{+/-}$, or NC1p with neutron tagging. }
    \centering
    \begin{tabular}{c|c}
        Method and Selection & Integrated Efficiency  \\
         \hline
         Calorimetric NCinc          & 0.60 \\
         Calorimetric NC0$\pi^{+/-}$ & 0.47 \\
         Kinematic NCinc             & 0.45 \\
         Kinematic NC0$\pi^{+/-}$    & 0.37 \\
         Kinematic NC1p + n-tag      & 0.16 \\
    \end{tabular}
    \label{tab:eff}
\end{table}
In adopting a NC0$\pi^{+/-}$ selection, the efficiency of the calorimetric method is more significantly affected than that for the kinematic method as demonstrated by the integrated efficiencies for the different reconstruction methods and event selections, shown in Table~\ref{tab:eff}. The difference in efficiency is reduced to 10\%. 

Having demonstrated the benefits and limitations of our new reconstruction method, we now consider one possible use case for the kinematic method: searching for NC disappearance as a part of SBN's broader sterile neutrino search.


\section{Utilizing Neutral Current Interactions in Sterile Neutrino Searches}

Over the last few decades, a series of anomalous neutrino flavor oscillation measurements~\cite{lsnd}\cite{miniboone}\cite{reactor}\cite{gallium} have been made at short baselines ($L/E \sim$ 1~m/MeV) that could be explained by the existence of one or more eV-mass scale neutrinos that do not interact via any Standard Model process, so-called sterile neutrinos. To date, no long-baseline oscillation experiments have found evidence for sterile neutrinos~\cite{minossterile}\cite{icecubenull}\cite{novanull}. Global analyses~\cite{Dentler2018}\cite{Adamson2020}\cite{Kopp2013} have shown tension at the level of 4$\sigma$ between short-baseline $\nu_e$ appearance measurements and $\nu_e$/$\nu_\mu$ disappearance measurements. 

In the case of one eV-mass scale, stable sterile neutrino added to the Standard Model - known as a 3+1 model - at short baselines and hundred-MeV- to GeV-scale neutrino energies, before standard neutrino oscillations become significant, the disappearance and survival probabilities are described by Equation~\ref{eq:nudis} and \ref{eq:nuapp} respectively. In this regime, CP-violating phases can be ignored. This is not the case at long baselines (e.g. MINOS/MINOS+ and NO$\nu$A) as discussed in the next subsection. The mixing is dominated by $\Delta m^2_{41}$ and the last column of the 4x4 extended PMNS matrix elements, $U_{\alpha 4}$, where $\alpha$ refers to the neutrino flavors, $\alpha \in \{e,\mu,\tau,s\}$. Here, s refers to the new sterile flavor. The effective mixing angles are related to the 4x4 PMNS matrix elements as shown in Equations~\ref{eq:thetadis} and \ref{eq:thetaapp}. $\Delta m_{41}^2$ is the mass-squared difference associated with the new sterile mass state. $L$ and $E_\nu$ are the propagation distance and energy of the neutrino respectively. 

\begin{equation}
    \label{eq:nudis}
    P_{\nu_\alpha \rightarrow \nu_\alpha} \simeq 1-\sin^2( 2\theta_{\alpha\alpha})\sin^2\left(1.27\Delta m_{41}^2 \frac{L}{E_\nu}\right)
\end{equation}

\begin{equation}
    \label{eq:nuapp}
    P_{\nu_\alpha \rightarrow \nu_\beta} \simeq \sin^2( 2\theta_{\alpha\beta})\sin^2\left(1.27\Delta m_{41}^2 \frac{L}{E_\nu}\right), \hspace{3pt} \alpha \neq \beta
\end{equation}

\begin{equation}
    \label{eq:thetadis}
    \sin^2(2\theta_{\alpha\alpha}) \equiv 4|U_{\alpha 4}|^2(1-|U_{\alpha 4}|^2)
\end{equation}

\begin{equation}
    \label{eq:thetaapp}
    \sin^2(2\theta_{\alpha\beta}) \equiv 4|U_{\alpha 4} U_{\beta 4}|^2, \hspace{3pt} \alpha \neq \beta
\end{equation}

For readers who are more accustomed to the parameterization more broadly used in oscillation experiments, we also provide the relations between $\theta_{\alpha\beta}$ and $\theta_{ij}$, $i,j=1,2,3,4$, for mixing angles that are relevant for this discussion.
\begin{equation}
   \begin{gathered}
    \label{eq:angletranslate}
        \sin^2 2\theta_{\mu\mu} = \sin^2 2 \theta_{24} \cos^2\theta_{14}+\sin^4 \theta_{24} \sin^2 2\theta_{14} \\
        \sin^2 2\theta_{\mu e} = \sin^2 2\theta_{14} \sin^2\theta_{24} \\
        \sin^2 2\theta_{\mu \tau} = \cos^4\theta_{14}\sin^22\theta_{24}\sin^2\theta_{34} \\
        \sin^2 2\theta_{\mu s} = \cos^4\theta_{14}\sin^22\theta_{24}\cos^2\theta_{34}
    \end{gathered}
\end{equation}

Motivated by the need for a definitive resolution to the short-baseline anomalies, the Short-Baseline Neutrino Program (SBN), hosted at Fermilab, was proposed in 2015~\cite{acciarri2015proposal}. SBN consists of three, hundred-ton scale liquid argon time projection chambers (LAr TPCs) located along the Booster Neutrino Beam (BNB) axis at distances of hundreds of meters from the BNB target: a near detector, SBND; an intermediate detector, MicroBooNE; and a far detector, ICARUS. Table~\ref{tab:sbndets} summarizes the detector masses and positions. The detector positions are optimized for sterile neutrino induced oscillations with a mass-squared difference of order 1~eV$^{2}$. SBN sensitivity studies have focused on $\nu_{\mu} (\bar{\nu}_{\mu})$ disappearance and $\nu_e(\bar{\nu}_{e})$ appearance, identified via charged current (CC) interactions. The most recent sensitivity study~\cite{machado2019} shows that SBN is positioned to cover the LSND 90\% C.L. allowed region and most of the globally allowed regions with 5$\sigma$ significance.

\begin{table}[tbh]
    \centering
        \caption{SBN detector parameters used in this analysis. All values are taken from~\cite{acciarri2015proposal}.}
    \begin{tabular}{c|c|c|c}
        Detector & Active Mass  & BNB Target   & Exposure\\
                 & [tons]       & Distance [m] & $\times10^{20}$[POT] \\
        \hline
         SBND       & 112 & 110 & 6.6 \\
         MicroBooNE & 89  & 470 & 13.2\\
         ICARUS     & 476 & 600 & 6.6 \\
    \end{tabular}

    \label{tab:sbndets}
\end{table}

A study evaluating SBN's sensitivities over a larger range of models than the minimal 3+1 model, including 3+2 and 3+3 with a consideration of CP-violating phases, can be found in~\cite{PhysRevD.96.055001}. There, as in the SBN proposal, only CC interactions were considered. 

We present a statistics only sensitivity analysis in the context of a 3+1 model, a commonly used benchmark in sterile neutrino search sensitivity studies. Note that sensitivities in the 3+1 case roughly translate into sensitivities in a 3+N scenario.

\begin{figure}[htb]
    \centering
    \includegraphics[width=\linewidth]{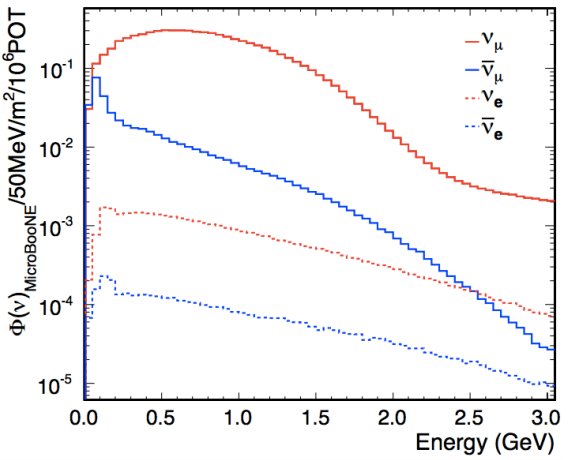}
    \caption{The BNB flux prediction at the SBN intermediate detector (MicroBooNE)~\cite{acciarri2015proposal} is shown for the neutrino running configuration (forward horn current).}
    \label{fig:flux}
\end{figure}
As shown in Figure~\ref{fig:flux}, the BNB flux peaks at about 600~MeV. The flavor content is about 93.6\% $\nu_\mu$ and 5.9\% $\bar{\nu}_\mu$ with contamination from $\nu_e$ and $\bar{\nu}_e$ at the level of 0.5\% below energies of 1.5~GeV. Given the low level of contamination from  $\nu_e$'s and $\bar{\nu}_e$'s, their contribution to the NC disappearance signal will be neglected in our analysis. This flux is produced with the BNB forward horn current configuration. Note that the BNB can also be run in a $\bar{\nu}_\mu$-dominant mode by inverting the focusing horn polarity (reverse horn current). The reverse horn current configuration is not considered in this analysis.

In a sterile neutrino search based on NC interactions, the signal is the disappearance of any active neutrinos. NC disappearance is an interesting channel as it is agnostic to active flavors; NC disappearance is sterile neutrino appearance. NC disappearance, quantified  by $1 - P_{\nu_\mu \rightarrow \nu_s}$, provides the only means of directly constraining $|U_{s4}|$. Note that the disappearance probability is the same for $\bar{\nu}_\mu$'s. The corresponding effective mixing angle, $\sin^2 2\theta_{\mu s}$, can be related to other mixing angles being probed at SBN by imposing unitarity on the 4x4 PMNS matrix (Equation~\ref{eq:unitary}), yielding the relation shown in Equation~\ref{eq:musrelate}, providing a link between the $\nu_\mu$/$\bar{\nu}_\mu$ disappearance, $\nu_\mu\rightarrow\nu_e$/$\bar{\nu}_\mu\rightarrow \bar{\nu}_e$ appearance, and $\nu_\mu\rightarrow\nu_\tau$/$\bar{\nu}_\mu\rightarrow \bar{\nu}_\tau$ appearance mixing angles.

\begin{equation}
    \label{eq:unitary}
    1 = \sum_{i=e,\mu,\tau,s} |U_{i4}|^2   
\end{equation}
\begin{equation}
    \label{eq:musrelate}
    \sin^2 2\theta_{\mu s} = \sin^2 2\theta_{\mu\mu} - \sin^2 2\theta_{\mu e} - \sin^2 2\theta_{\mu\tau}
\end{equation}

\subsection{Constraints on PMNS Matrix Elements}
Equation~\ref{eq:musrelate} demonstrates the additional physics reach that could be provided to SBN by the addition of a NC disappearance search. In addition to providing complimentary information to the SBN $\nu_{\mu}$ disappearance and $\nu_e$ appearance analyses, a NC disappearance search at SBN can provide unique constraints on the 3+1 phase space.

Past sterile neutrino searches have contributed to a growing collection of data sets, providing constraints on $|U_{e4}|$ and $|U_{\mu 4}|$. There are relatively few constraints on $|U_{\tau 4}|$ due to the lack of a $\nu_\tau$ source and the few-GeV energy threshold for $\nu_\tau$ CC interactions, nearly beyond the reach of most neutrino sources. To date, the most stringent constraint on $|U_{\tau 4}|$ comes from atmospheric neutrino oscillations observed in IceCube~\cite{collin2016}. The IceCube measurements utilize the MSW effect to probe $|U_{\tau 4}|$ via CC interactions.  

We are aware of only two experiments that have set limits in the case of a sterile neutrino NC disappearance search, NO$\nu$A~\cite{nova_thesis}\cite{nova_ncdis} and MINOS/MINOS+~\cite{minos_ncdis}. These two experiments and IceCube each set mutually consistent limits in the $(|U_{\mu4}|,|U_{\tau 4}|)$ plane with the IceCube limit being more stringent.

Since MINOS/MINOS+ and NO$\nu$A operate at long baselines, the NC disappearance probability dependence on $\Delta m_{31}^2$ becomes important for describing oscillations at the far detector, introducing more free parameters into the oscillation fit. As is shown in Equation~\ref{eq:lbncdis}, there are two corrections to the short-baseline limit that are driven by $\Delta m_{31}^2$. These additional terms introduce two additional parameters to the probability calculation: one mixing angle from standard oscillations, $\theta_{23}$ and one new CP-violating phase, $\delta_{24}$. These challenges were addressed by incorporating the near detector into a simultaneous two-detector fit, for example see a recent result from MINOS+~\cite{minossterile}. Oscillations occurring in the near detector are still well described in the short-baseline limit. Such measurements are complimentary to SBN, where all three detectors operate in the short-baseline limit. With a NC disappearance analysis, SBN should be able provide one of the most stringent constraints on $|U_{\tau 4}|$ to date.

\begin{equation}
\begin{gathered}
1-P_{\nu_\mu \rightarrow \nu_s} \approx \\
1 - \cos^4\theta_{14} \cos^2\theta_{34}\sin^2 2\theta_{24} \sin^2 \left ( \frac{\Delta m_{41}^2 L}{4E}\right ) \\
- \sin^2\theta_{34} \sin^2 2\theta_{23} \sin^2 \left ( \frac{\Delta m_{31}^2 L}{4E} \right ) \\
+ \frac{1}{2} \sin \delta_{24} \sin \theta_{24} \sin 2\theta_{23} \sin \left ( \frac{\Delta m^2_{31} L}{2E} \right )
\label{eq:lbncdis}
\end{gathered}
\end{equation}

In evaluating the impact of a NC analysis at SBN, it is useful to compare our sensitivities to the globally allowed phase space. However, this is difficult in that, to our knowledge, a global analysis of sterile neutrino induced NC disappearance does not exist. In order to provide some baseline with which we can benchmark our performance, we used the globally allowed values of the PMNS matrix elements, specified for two values of $\Delta m_{41}^2$, found in~\cite{collin2016}. This supplies four points in the $(\sin^22\theta_{\mu s},\Delta m_{41}^2)$ plane after applying Equations~\ref{eq:thetadis}, \ref{eq:thetaapp}, and \ref{eq:musrelate}. While the allowed region that this procedure produces is likely not complete, it does provide a reasonable set of values to which we can compare limits that we produce. Global analyses of sterile neutrino searches and likewise calculating globally allowed regions in phase space is difficult. For more discussion, \cite{DIAZ20201} gives a summary of the current status and challenges of sterile neutrino searches.

\section{Applying Neutral Current Reconstruction at SBN}

We have established that the conventional calorimetric method is better in terms of efficiency while our new kinematic method performs better in accurately and precisely reconstructing the neutrino energy. In this section, we apply each reconstruction method to the search for sterile neutrinos at SBN via a NC disappearance search in the context of the minimal 3+1 model.

We demonstrate that the calorimetric and kinematic methods are each sufficient for observing spectral features that are characteristic of neutrino disappearance. This is important not only for observing disappearance but also for determining the values of the mixing parameters, $\Delta m^2_{41}$ and $\sin^22\theta_{\mu s}$. We present a comparison of the true and reconstructed neutrino energy spectra at each SBN-like detector. Next, we quantify our sensitivity to eV-scale sterile neutrinos in the context of the minimal 3+1 model. We show exclusion limits followed by allowed regions for selected nonzero mixing parameters.

For this analysis, each of the detector specific samples is scaled to the expected exposure provided in the SBN proposal: 6.6~x~10$^{20}$~POT for the near and far detectors, 13.2~x~10$^{20}$~POT for the intermediate detector. We consider only exclusive NC0$\pi^{+/-}$ samples, identified by the reconstruction, for this sensitivity analysis. $\nu_\mu$CC events where the final state muon and any charged pions have momenta blow the tracking threshold contribute to event rates at the percent-level so are neglected from this analysis. We do anticipate some background from beam induced activity in materials surrounding the argon active volume, so-called dirt events. These are not expected to comprise a significant fraction of our NC events. However, all possible sources of background will be addressed in followup studies using a full detector simulation and reconstruction.

We present sensitivities for the calorimetric method with NC0$\pi^{+/-}$ selection, the kinematic method with NC0$\pi^{+/-}$ selection, and the kinematic method with NC1p and n-tagging selection. For these sensitivities, we only present results obtained without Monte Carlo corrections to the reconstructed neutrino energy. We verified that the sensitivities are not affected by the introduction of such corrections.

\subsection{Neutrino Energy Spectra}
\begin{figure*}[p]
  \centering
  \begin{subfigure}{0.78\textwidth}  
    \includegraphics[width=\linewidth]{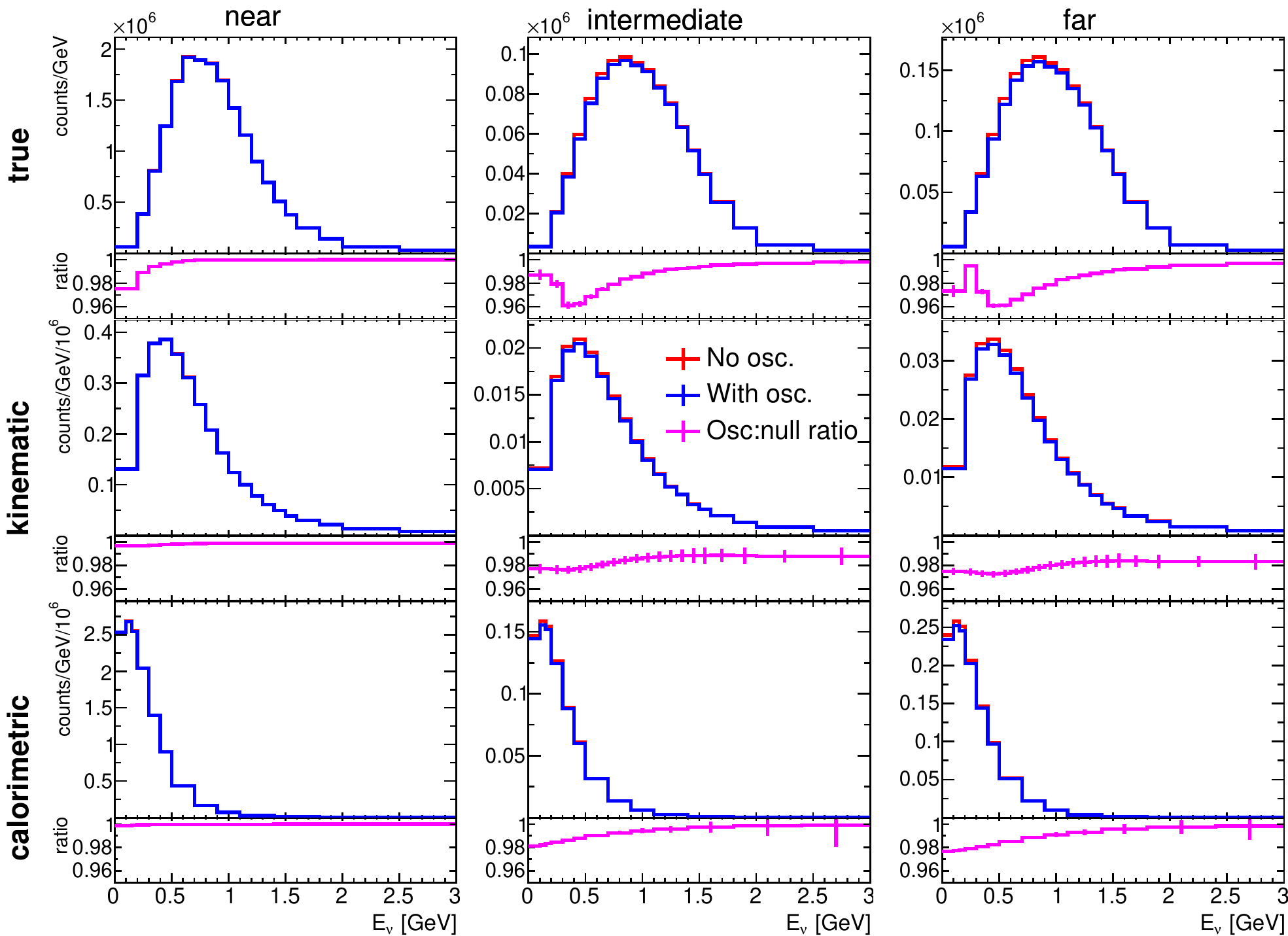}
    \subcaption{$\Delta m^2_{41}$ = 1~eV$^2$, $\sin^2 2\theta_{\mu s}$ = 0.04}
    \label{fig:spectra_low}
  \end{subfigure}
  \begin{subfigure}{0.78\textwidth}  
    \includegraphics[width=\linewidth]{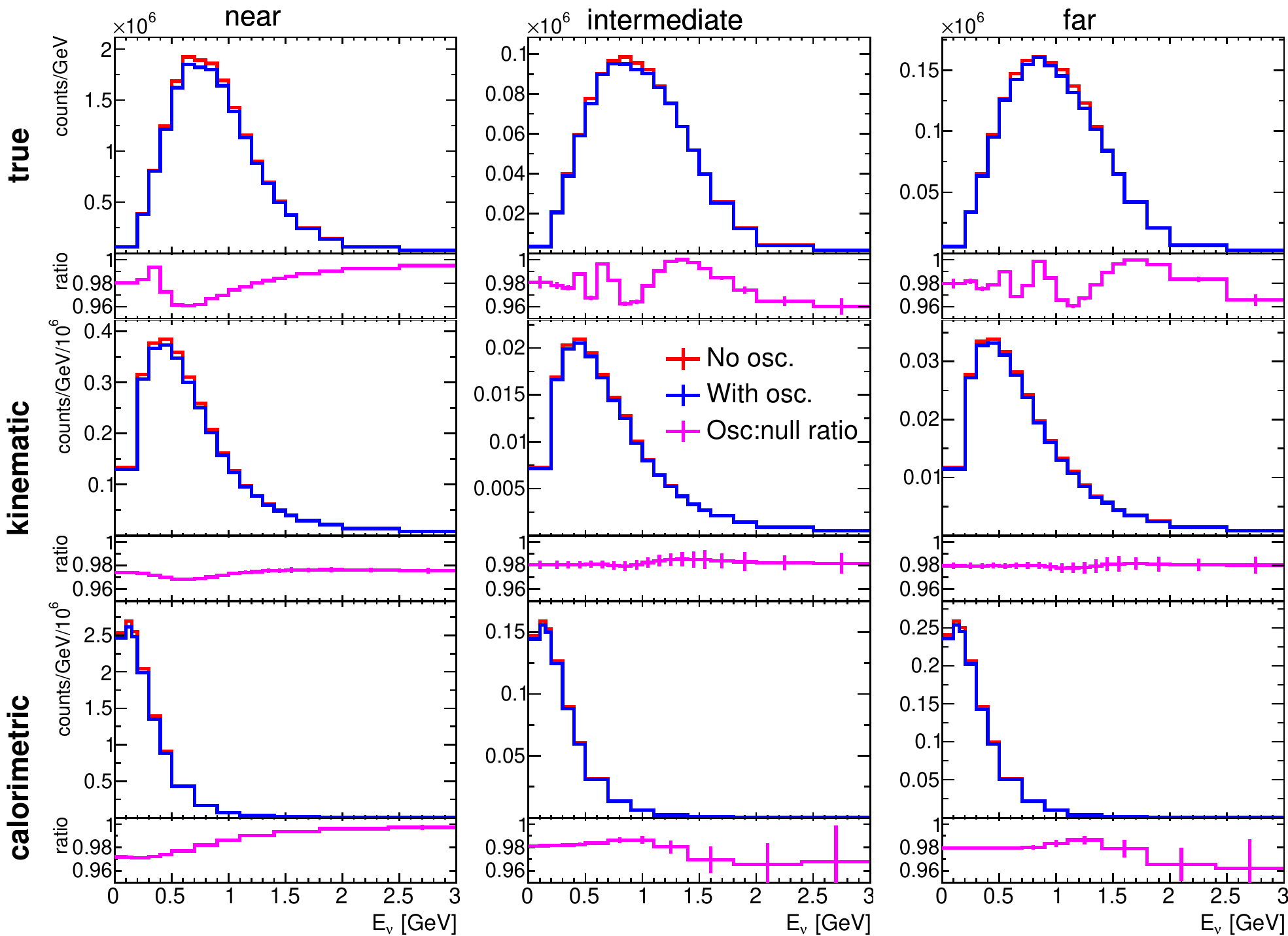}
    \subcaption{$\Delta m^2_{41}$ = 7~eV$^2$, $\sin^2 2\theta_{\mu s}$ = 0.04}
    \label{fig:spectra_high}
  \end{subfigure}
  \caption{NC neutrino energy spectra with and without oscillations are shown for each SBN-like detector: near (left column), intermediate (center column), and far (right column). There is one row per method: true (top row), kinematic NC1p + n-tag (center row), and calorimetric NC inclusive (bottom row).}
  \label{fig:spectra}
\end{figure*}

To illustrate the power of resolving spectral features in observing oscillation effects, Figure~\ref{fig:spectra} shows NC neutrino energy spectra for each SBN-like detector. The null hypothesis, absence of short-baseline oscillations, is compared to a case with $\sin^2 2\theta_{\mu s} = 0.0403$ and $\Delta m^2_{41} = $ 1~eV$^2$ (\ref{fig:spectra_low}) or $\Delta m^2_{41} = $ 7~eV$^2$ (\ref{fig:spectra_high}). These values are taken from the edge of the globally allowed region in the 3+1 phase space.

The oscillation probabilities are evaluated at a single, fixed baseline for each detector, set by the detector distance to the BNB target (see Table~\ref{tab:sbndets}). In reality, the precise production point of the neutrino is smeared out by the length of the decay pipe, introducing a systematic uncertainty that will cause smearing of the oscillation features. This will be addressed in a future study. 

The $L/E_\nu$ behavior of the oscillations is evident. A clear, energy dependent depletion in events is observable in the near detector for both values of $\Delta m^2_{41}$ while the energy dependent depletion at the intermediate and far detectors is primarily only resolvable at lower values.

Comparing the calorimetric and kinematic methods, we see that the kinematic method performs better in pinpointing the location of the oscillation maximum. While the calorimetric method yields some shape information, it does not directly provide the oscillation peak location, however it has the sizeable advantage of significantly higher statistics. Comparing the low- and high-$\Delta m^2_{41}$ cases shows how both methods have sensitivity to the true energy spectral distortions.

In the following subsection, we describe how the advantages of both methods may be combined to boost the oscillation sensitivity.

\subsection{Oscillation Sensitivity}

In order to evaluate the sensitivity of our method in the context of SBN, we adopt a binned Poisson likelihood approach to evaluate the statistical significance of a 3+1 signal compared to the null hypothesis with no sterile neutrinos. We use the Asimov data set and construct a log likelihood ratio ($\mathcal{L}$) as shown in Equation~\ref{eq:llr}, where $N^{exp}$ is the expected number of events given some test hypothesis, and $N^{obs}$ is the observed number of events. Assuming Wilk's Theorem applies, we can obtain the statistical significance $\sigma = \sqrt{\Delta \chi ^2} = \sqrt{\mathcal{L}}$.   $\Delta \chi^2$ is the $\chi^2$ difference from the global minimum, corresponding to the maximum likelihood that is achieved with the Asimov data set. We use an approximate form of $\mathcal{L}$ obtained through the application of Sterling's approximation for sensitivity calculations, shown in Equation~\ref{eq:llrapprox}. The index runs over bins of reconstructed neutrino energy across all three SBN-like detectors. For consistency, we verified that our analysis methods produce a similar result for the 3+1 statistics-only $\nu_\mu$ disappearance sensitivity analysis as in~\cite{PhysRevD.96.055001}.

\begin{equation}
    \label{eq:llr}
    \mathcal{L} = -2\ln \left [ \prod_{i}^{N}  \frac{(N_i^{obs})^{N_i^{exp}}}{N_i^{exp}!} \text{e}^{-N_i^{obs}}  \right]
\end{equation}

\begin{equation}
    \label{eq:llrapprox}
    \mathcal{L} \approx 2\sum_{i}^{N} \left [ N_i^{obs} - N_i^{exp} + N_i^{exp} \text{ln}\frac{N_i^{exp}}{N_i^{obs}}  \right]
\end{equation}

\begin{figure}[htb]
    \centering
    \includegraphics[width=\linewidth]{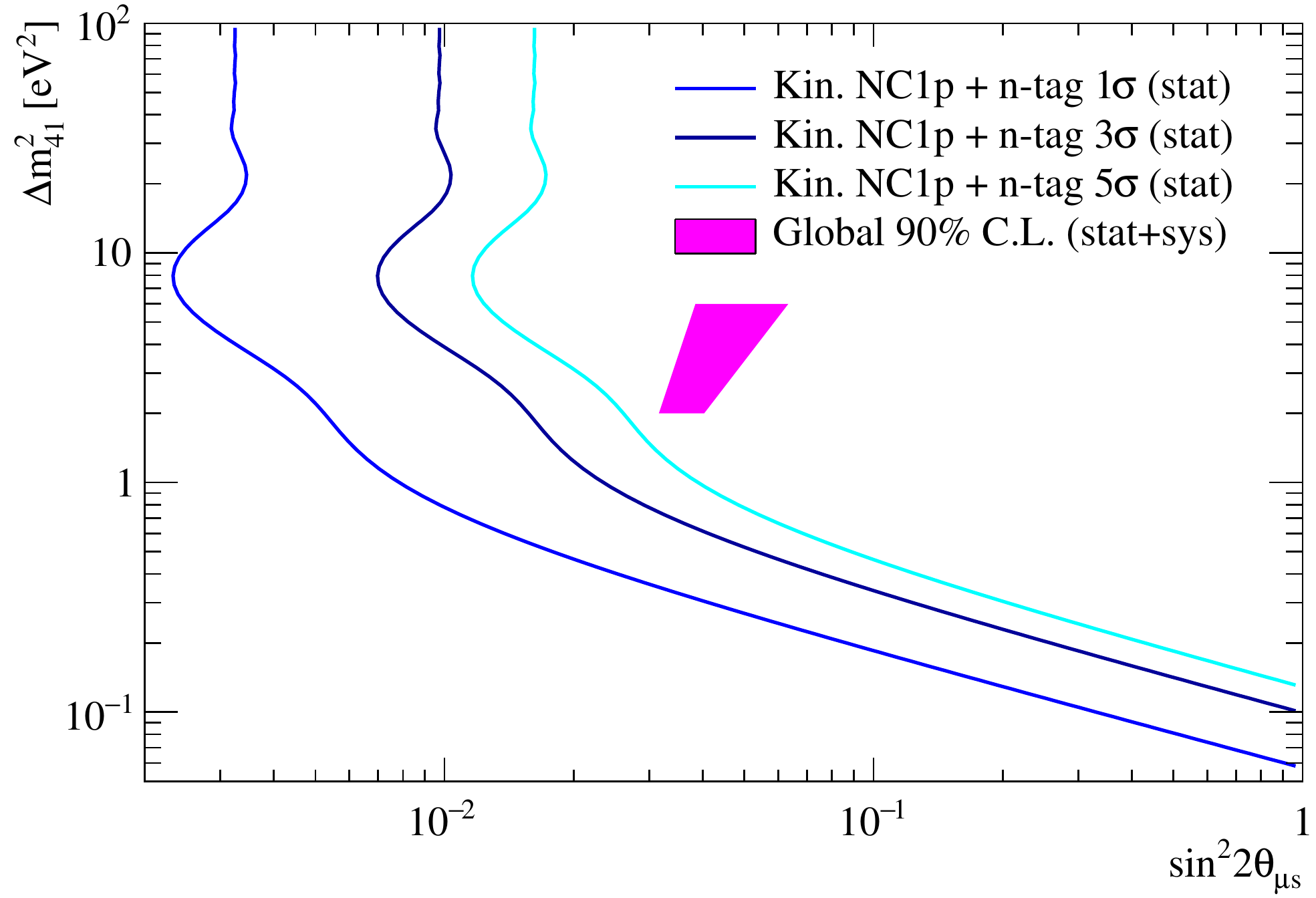}
    \caption{Using the kinematic reconstruction method with NC1p and n-tagging selection, exclusion contours are shown for a SBN-like, three-detector fit including statistical uncertainties only. 1$\sigma$, 3$\sigma$, and 5$\sigma$ contours are shown along with the globally allowed 90\% C.L. that includes both statistical and systematic uncertainties~\cite{collin2016}.}
    \label{fig:oscsens}
\end{figure}

The neutral current disappearance sensitivity for the kinematic method using a NC1p with n-tagging selection is shown in Figure~\ref{fig:oscsens} with 1$\sigma$, 3$\sigma$, and 5$\sigma$ contours drawn in $(\sin^22\theta_{\mu s},\Delta m^2_{41})$ space. The strength of the limit increases as $\Delta m^2_{41}$ increases due to the oscillation maximum moving closer to the far then intermediate detectors. The enhancement at $\Delta m^2_{41} \sim$~8~eV$^2$ is where the oscillation maximum occurs in the near detector.

The contours in this analysis include statistical uncertainties only. The addition of systematic uncertainties is expected to reduce sensitivity primarily for $\Delta m^2_{41} >$ 1~eV$^2$ where the oscillation probability begins to vary rapidly with neutrino energy. For comparison, globally allowed values from ~\cite{collin2016}, which include statistical and systematic uncertainties, are shown to be covered with $5\sigma$ significance. 

\begin{figure}[htb]
    \centering
    \includegraphics[width=\linewidth]{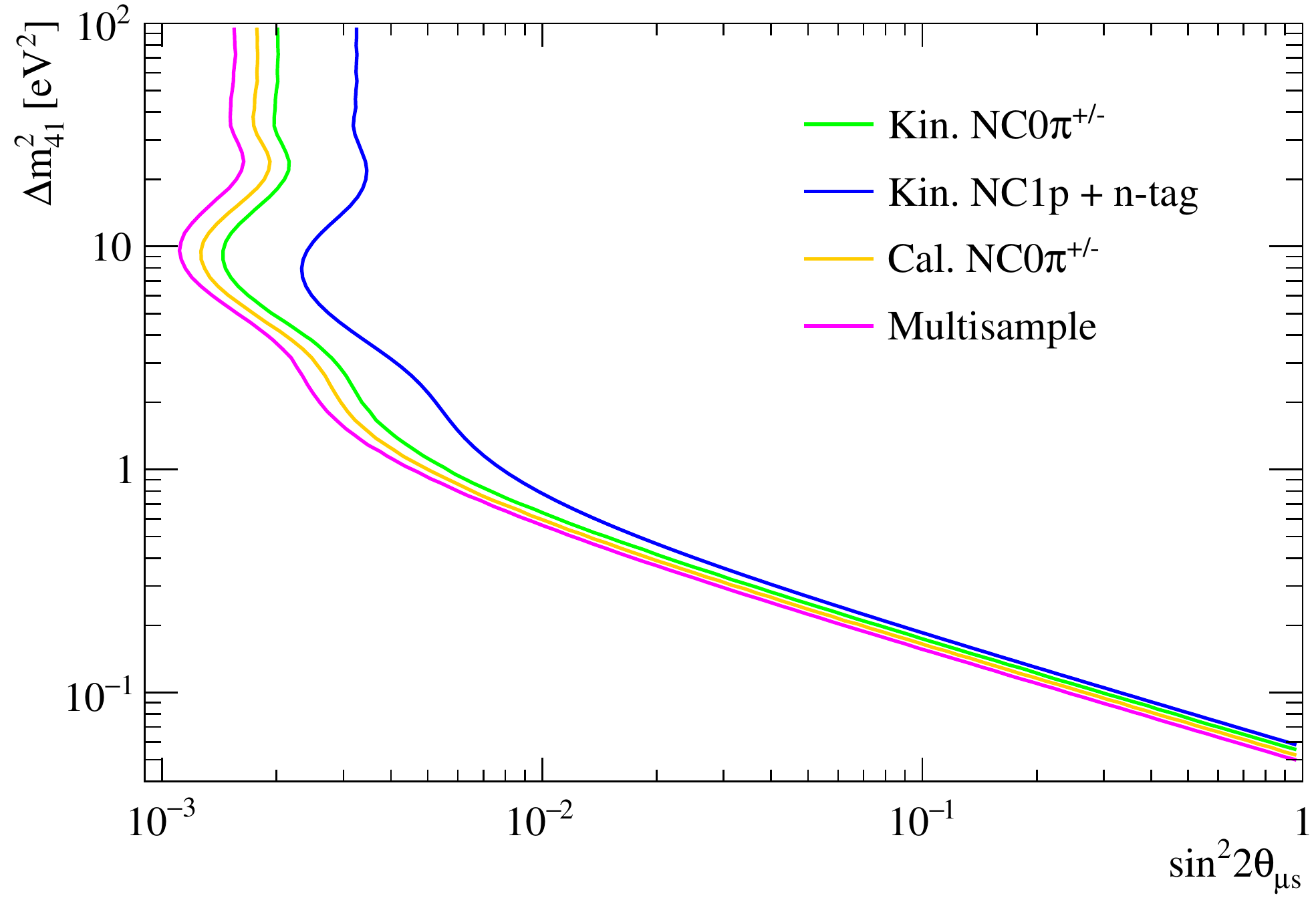}
    \caption{1$\sigma$ exclusion contours are shown for a SBN-like, three-detector fit including statistical uncertainties only for the different reconstruction methods and sample selections.}
    \label{fig:senscompare}
\end{figure}
While our kinematic reconstruction method is sufficient to cover the globally allowed region with 5$\sigma$ significance, its limited efficiency yields a weaker NC disappearance sensitivity compared to the calorimetric method. In setting limits in our case, statistics are the limiting factor. This motivates a multi-sample approach where we combine the kinematic and calorimetric methods - we want to extract the more accurate and precise shape information with the kinematic method while maintaining the same reconstruction efficiency as the calorimetric method. 

For the multi-sample approach, we construct three statistically independent samples of reconstructed neutrino energy using a combination of reconstruction methods and event selections: kinematic NC1p with n-tagging, kinematic NC0$\pi^{+/-}$, and calorimetric NC0$\pi^{+/-}$. First, the kinematic method with a NC1p and n-tagging selection is applied. Any events rejected by the selection are passed to the kinematic method with a NC0$\pi^{+/-}$ selection. Finally, any events rejected in this step, those with a negative reconstructed energy, are passed to the calorimetric method with a NC0$\pi^{+/-}$ selection.
 
1$\sigma$ exclusion limits for the four different methods and selections are shown in Figure~\ref{fig:senscompare}. The statistical power of the calorimetric method is evident compared to kinematic method selections. In the multi-sample selection that maintains the same overall sample size as the calorimetric method, the added shape information from the kinematic method improves the sensitivity by approximately 10\%.

\begin{figure}[htb]
    \centering
    \includegraphics[width=\linewidth]{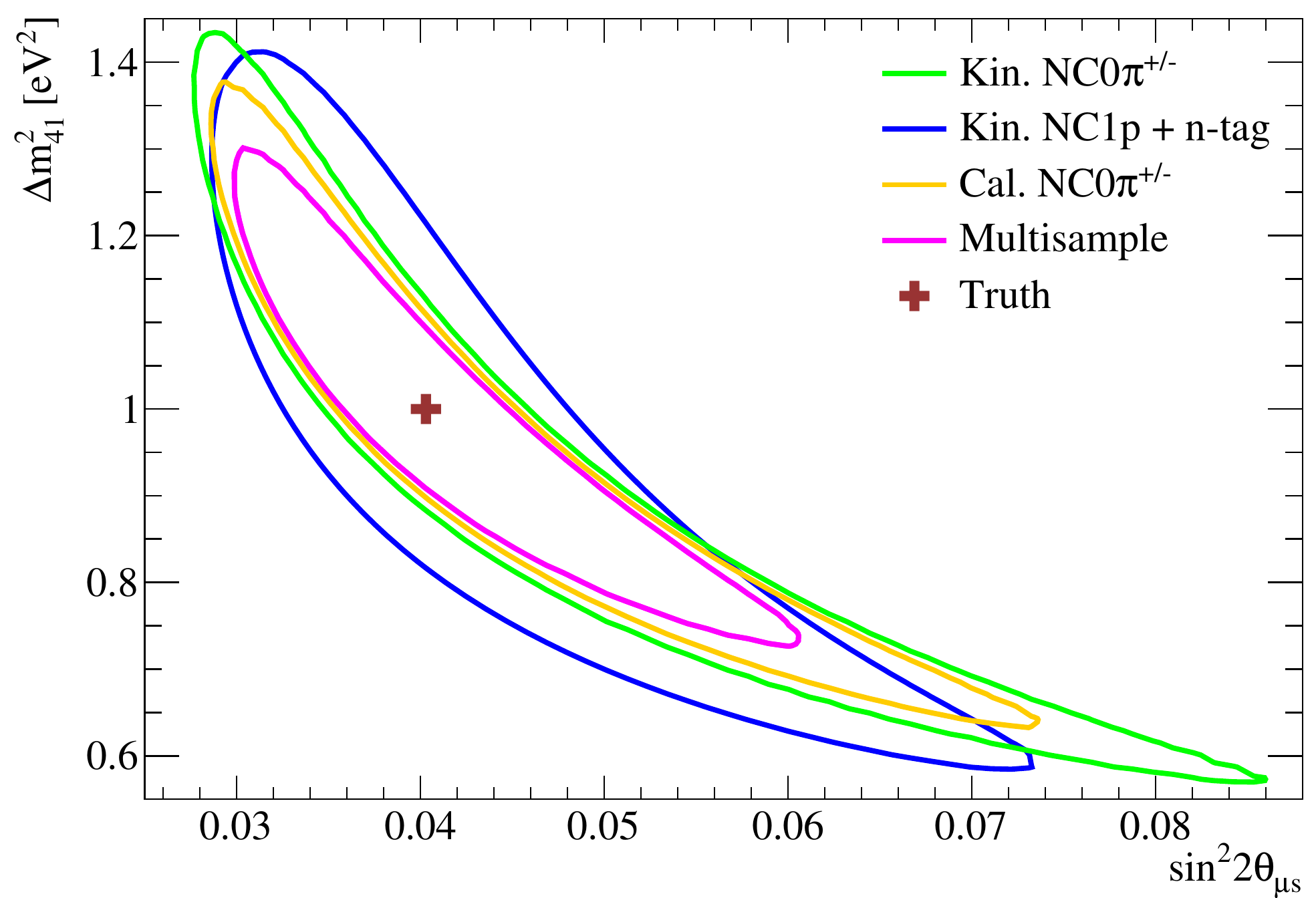}
    \caption{For the case with $\Delta m^2_{41}$ = 1~eV$^2$ and $\sin^2 2\theta_{\mu s}$~=~0.04, 1$\sigma$ allowed regions are shown for a SBN-like, three-detector fit including statistical uncertainties only for the different reconstruction methods and sample selections.}
    \label{fig:allowcompare_low}
\end{figure}
\begin{figure}[htb]
    \centering
    \includegraphics[width=\linewidth]{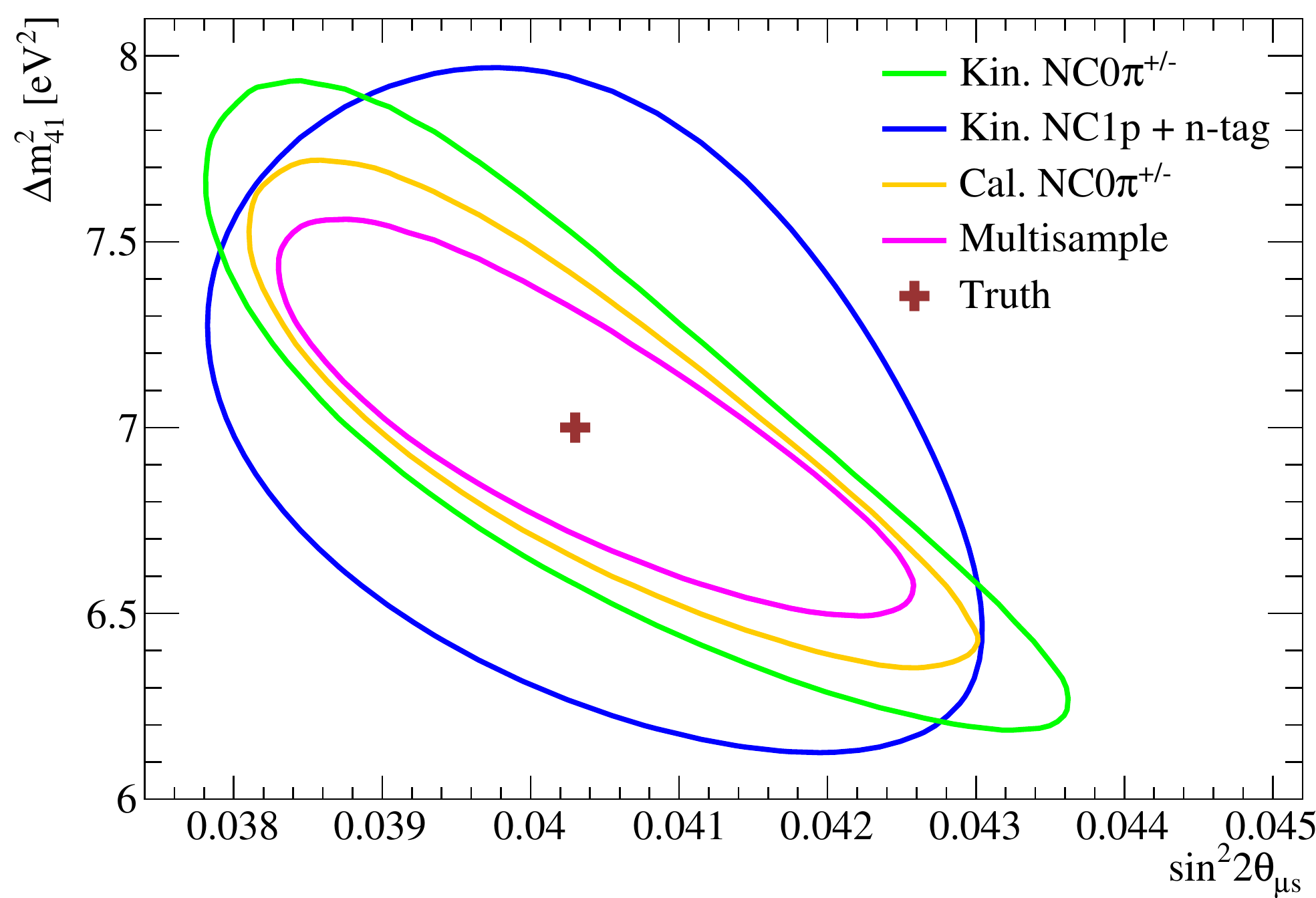}
    \caption{For the case with $\Delta m^2_{41}$~=~7~eV$^2$ and $\sin^2 2\theta_{\mu s}$~=~0.04, 1$\sigma$ allowed regions are shown for a SBN-like, three-detector fit including statistical uncertainties only for the different reconstruction methods and sample selections.}
    \label{fig:allowcompare_high}
\end{figure}
Note that it is possible to improve the limits shown in Figure~\ref{fig:senscompare} in some regions of phase space by increasing the sample size at the far detector. For larger values of $\Delta m^2_{41}$, the limits are expected to be systematics limited, driven by the shape measurement in the near detector. For smaller $\Delta m^2_{41}$, however, the limits are constrained by the far detector statistics. For example, a doubling of the planned exposure at the far detector results in an approximate 50\% sensitivity improvement for 0.5~eV$^2 < \Delta m^2_{41} < $~4~eV$^2$, where the oscillation maximum is in close proximity to the far detector.

The power of the multi-sample approach becomes more evident when considering the ability to pinpoint the oscillation parameters in the presence of a disappearance signal where shape information is crucial. Figures~\ref{fig:allowcompare_low} and \ref{fig:allowcompare_high} show the 1$\sigma$ allowed regions for the same mixing parameter values shown in Figures~\ref{fig:spectra_low} and \ref{fig:spectra_high}. In both cases, the multi-sample fit clearly outperforms the calorimetric method alone. For both cases, with $\sin^22\theta_{\mu s} = 0.0403$ and $\Delta m^2_{41} = 1~\text{eV}^2$ or $\Delta m^2_{41} = 7~\text{eV}^2$, the multi-sample approach produces a 30\% smaller allowed region compared to the calorimetric method with NC0$\pi^{+/-}$ selection. 

Additionally, we expect the use of two different energy estimation methods will provide a way of cross-checking biases from interaction modelling uncertainties, reducing the final systematic uncertainty. While the selected distributions are not identical, there is substantial overlap.

\section{Conclusions and Outlook}

Estimating the incident neutrino energy in neutral current interactions is challenging. The conventional method of relying on calorimetry alone, measuring the total visible energy, provides a lower bound on the neutrino energy, but the accuracy and precision is poor relative to methods used in charged current interactions. The excellent calorimetry and tracking capabilities and low tracking thresholds of liquid argon time projection chambers (LAr TPCs) motivate a new method for reconstructing the incoming neutrino energy based on the kinematics of the final state hadronic system. 

While our new kinematic method provides significantly better energy resolution and reduced bias compared to the conventional calorimetric method, it suffers from low efficiency. In combining our method with the conventional one in a multi-method, multi-sample approach, we showed that we can push oscillation sensitivities beyond what is possible with the conventional method alone.

Our toy study illustrates how neutral current interactions can be utilized in the context of sterile neutrino searches, the 3+1 model in particular. As a relevant example of how our approach can be applied, we showed that the addition of a neutral current disappearance search to the Short-Baseline Neutrino Program (SBN) can potentially enhance its physics reach, providing both complimentary and redundant information that can serve to overconstrain the sterile neutrino phase space in the currently considered $\nu_\mu$ disappearance and $\nu_e$ appearance channels. In addition, our approach could set unique limits on the poorly constrained 3+1 PMNS matrix element, $U_{\tau 4}$.

We emphasize that this study is a toy study with several caveats. First, we adopt optimistic reconstruction performance parameters. Next, we assume we can reduce charged current backgrounds to negligible levels by rejecting any neutral current event candidates with a muon-like or charged pion-like track. We further assume that other sources of background are negligible. In addition, our study neglects the distribution of neutrino production points in the decay pipe, an effect that smears out spectral oscillation features. Finally, we do not account for systematic uncertainties. These points need to be addressed in a more thorough, detailed analysis.

With this proof of concept, we will implement our neutral current analysis approach in the SBN simulation and reconstruction framework. We will test our method using a full simulation and reconstruction and reevaluate the performances presented in this work. Finally, the neutral current analysis will be integrated into the oscillation sensitivity analysis framework to provide the most realistic sensitivity projections.

While we wait for SBN to accumulate data, we can take advantage of the data set obtained by MicroBooNE to test our reconstruction method. This is a crucial step as the performance we have presented here is model dependent. For example, if nuclear effects are more or less significant, this will have an impact on the energy resolution and selection efficiency of the neutral current analysis and, therefore, the sensitivities presented in this work. To this end, we can make use of charged current interactions, replacing the outgoing neutrino in our model with the outgoing lepton. This will provide a reliable test of using the kinematics of the final state hadronic system as an estimator of the incoming neutrino energy.

Looking further into the future, the Deep Underground Neutrino Experiment (DUNE) near detector will be exposed to what will be the world's most intense neutrino beam. Along with state-of-the-art detectors, including a LAr TPC and a magnetized high pressure gaseous argon TPC, the performance of our neutral current analysis approach should be improved. In a future study, we will investigate the application of our approach to neutral current inclusive and exclusive cross section measurements at DUNE.

\begin{acknowledgments}
This work was supported by funding from the University of Minnesota.
\end{acknowledgments}

\nocite{*}

\bibliography{references}

\end{document}